\documentclass[11pt]{article}

\usepackage{natbib}
\usepackage{graphicx}
\usepackage{amsmath,amssymb,amsthm}
\usepackage{array,multirow}
\usepackage[colorlinks=false,linkcolor=blue,citecolor=blue,urlcolor=blue,bookmarks=false]{hyperref}
\usepackage{url}
\usepackage{verbatim}
\usepackage[top=1in, bottom=1in, left=1in, right=1in]{geometry}

\usepackage{caption}
\usepackage[justification = centering]{subcaption}
\usepackage{lscape}

\usepackage{setspace}

\def\b1{\boldsymbol{1}}
\newcommand{\ind}[1]{\b1_{\left\{#1\right\}}}

\newcommand{\pr}[1]{\mathbf{P} \! \left \{ #1 \right \}}
\newcommand{\EE}[1]{\E \left [ #1 \right ]}

\usepackage{booktabs}
\usepackage[flushleft]{threeparttable}

\newcommand{\E}{\mathbb{E}}

\usepackage{pifont}

\theoremstyle{definition}
\newtheorem{assump}{Assumption}
\newenvironment{myassump}[2][]
  {\renewcommand\theassump{#2}\begin{assump}[#1]}
  {\end{assump}}

\usepackage{color}

\begin{document}

\title{Bounding, \\an accessible method for estimating principal causal effects,\\ examined and explained\thanks{Luke Miratrix (\url{luke_miratrix@gse.harvard.edu})is the corresponding author.  We gratefully acknowledge financial support from the Spencer Foundation through a grant entitled ``Using Emerging Methods with Existing Data from Multi-site Trials to Learn About and From Variation in Educational Program Effects,'' and from the Institute for Education Science (IES Grant \#R305D150040). We thank Julie Edmunds and the North Carolina Education Research Data Center for facilitating access to the ECHS data and Julie Edmunds and Fatih Unlu for feedback and guidance on the ECHS data and study context. This publication uses data collected by the SERVE Center at UNCG as supported by the Institute of Education Sciences under grants numbered \#R305R060022 and \#R305A110064. We thank Ed Bein and Howard Bloom for early comments on the draft. Any opinions, findings, and conclusions or recommendations expressed in this publication are those of the authors and do not necessarily reflect the views of the SERVE Center, the Spencer Foundation, or the Institute of Education Sciences.
}}

\date{\today}

\author{Luke Miratrix\\Harvard University  \and Jane Furey\\Bundesinstitut f\"ur Berufsbildung \and Avi Feller\\UC Berkeley\and Todd Grindal\\ SRI International \and Lindsay C. Page\\University of Pittsburgh  }
\date{\today}

\maketitle

\begin{abstract}
Estimating treatment effects for subgroups defined by post-treatment behavior (i.e., estimating causal effects in a principal stratification framework) can be technically challenging and heavily reliant on strong assumptions. We investigate an alternative path: using bounds to identify ranges of possible effects that are consistent with the data. This simple approach relies on fewer assumptions and yet can result in policy-relevant findings. As we show, covariates can be used to substantially tighten bounds in a straightforward manner. Via simulation, we demonstrate which types of covariates are maximally beneficial. We conclude with an analysis of a multi-site experimental study of Early College High Schools. When examining the program's impact on students completing the ninth grade ``on-track'' for college, we find little impact for ECHS students who would otherwise attend a high quality high school, but substantial effects for those who would not. This suggests potential benefit in expanding these programs in areas primarily served by lower quality schools. 
\end{abstract}

\vspace{3mm}

\section{Introduction}

With the proliferation of randomized trials in education, researchers are asking ever more sophisticated questions about program impacts \citep{spybrook2014detecting}. Collectively, the field is evolving from first-order questions about ``what works overall'' to more nuanced questions about what works, for whom, when, and under what circumstances \citep{raudenbush2015learning}. That is, researchers and policy makers are interested in better understanding the many ways that impacts may vary across contexts and subpopulations.  When relevant groups are defined by observed, pre-randomization characteristics, the process for generating causal estimates within subgroups is typically straightforward.
Yet key questions often pertain to subgroups defined by behaviors, actions, or decisions that occur after randomization.

Principal stratification \citep{Frangakis:2002} provides a framework for considering such questions.
The first step is to specify subgroups, or \emph{principal strata}, defined by the combination of experimental subjects' observed and counterfactual post-randomization actions, behaviors or responses, and then to articulate estimands associated with each stratum.
A key concept of this approach is that because these subgroups are defined by the combination of potential actions subjects would take under the different experimental conditions, a subject's stratum membership is considered a pre-treatment characteristic, even though it may be unobserved. Within these strata, researchers then typically aim to estimate the \emph{principal causal effects}, which are the stratum-specific intent-to-treat effects of randomization.
Procedurally, the analyst first defines the principal strata and their associated estimands of interest, such as the share of experimental subjects belonging to each stratum or the stratum-specific principal causal effects.
A common next step is to apply scientific or substantive knowledge (i.e., assumptions) to restrict or inform these quantities, as we illustrate below.\footnote{For example, we might argue that certain strata do not exist (akin to the monotonicity or ``no defiers'' assumption critical to instrumental variables estimation) or that the treatment has no effect on outcomes in other strata (normally called an exclusion restriction).}
Only after these steps does the analyst turn to estimation.

The challenge in estimation is that, even with such restrictions, many quantities of interest are only ``set'' rather than ``point'' identified, meaning that the observed data are entirely consistent with a range of possible stratum parameter values, rather than just a single point.
Many methods to estimation therefore rely on imposing arguably quite strong distributional and/or independence assumptions to make progress. These strategies, such as model-based principal stratification (e.g., \cite{FeGrMiPa2016}; \cite{frumento2012evaluating}; \cite{page2012principal}) and principal scores (e.g., \cite{jo2009use}; \cite{stuart2015assessing}) can be quite technical to implement and easily can be critiqued as yielding results that are highly sensitive to their identifying assumptions. Further, even under these assumptions, estimator performance can be quite poor \citep{Feller:2015ux}.

An alternative approach is instead to focus on the \emph{set} of possible parameter values, which are typically straightforward to obtain.
To do this we typically bound the parameter of interest in an interval by assessing which values are consistent with the observed characteristics of the data.
See \cite{manski2013public} for an overview of this approach.
Unfortunately, the range of possible parameter values within each stratum is often too wide to be substantively meaningful.
Therefore, in this paper we illustrate how to capitalize on baseline covariates to tighten these bounds.
In particular, our key contributions are that we carefully consider procedures for selecting baseline covariates and investigate the relative gain in precision from selecting covariates of different types.
For a more thorough, and non-technical, introduction to principal stratification itself see \citet{Page:2015km}.
See also \cite{Mealli:dc}  and \cite{Schochet:2014wa}.

There is a sizable literature on bounds in this setting. We highlight three papers that explicitly use covariates to tighten bounds.
First, \cite{Lee:2009tc} utilizes bounds to assess the extent to which the effect of Job Corps on total earnings was a function of increased employment, an increased wage rate, or both.  The analytic challenge Lee faces is that any improvement in wage rate can only be assessed for those who would be employed regardless of the intervention. \cite{Grilli:2008ih} use a bounds approach in the context of observational studies. \cite{Long:2013hs} provide some technical conditions under which certain covariates tighten bounds.
We extend this work by investigating how an analyst should best select and combine different baseline covariates to tighten bounds in practice.

We structure the remainder of the paper as follows.  We first introduce a multi-site, multi-cohort randomized controlled trial investigating the Early College High School (ECHS) model. This experimental evaluation motivates and serves as the substantive focus of this work. We then introduce our key research question and devise the principal stratification setup and associated estimands of interest. We illustrate how to directly estimate certain stratum-specific quantities and how to obtain bounds for those quantities that are not point identified. Finally, we discuss the use of covariates for tightening bounds and detail the relative gains to be made with different covariate choices, both via simulation and via an analysis of our motivating example.

Throughout, we aim to underscore bounding as an approach that is computationally straightforward and that can provide general insight for principal causal effects.
Importantly, the analytic strategies that we present are transparent and can be adapted and used by the majority of quantitative researchers in education using standard software.\footnote{R and Stata routines for this context are available upon request.  This code could be used as a template for other scenarios.}
We also include the relevant details of our derivations in the hopes that illustrating the process step by step will help others conduct similar analyses for their own principal stratification scenarios.
As we show, the insights produced from covariate-tightened bounds can be of use in real contexts for guiding policy and programmatic decisions.

\section{Early College High Schools}

High school graduation is often a minimum requirement for adult labor market success. Nonetheless, nearly one fifth of U.S. students do not attain a high school degree \citep{kena2015condition}. Furthermore, among students who graduate from high school, many begin college or join the workforce without the skills necessary to succeed \citep{sparks2013first}. Educators and policymakers have responded to these problems by developing new secondary school models. One such model is the Early College High School (ECHS).

The Early College High School (ECHS) Initiative began in 2002 with the goal of increasing high school graduation rates and college enrollment. ECHSs are typically autonomous small schools managed by local school districts in partnership with two- or four-year colleges. ECHS students generally begin taking college-level courses in ninth grade and have the opportunity to graduate high school with an associate's degree and/or two years of transferable college credit.
There are three primary mechanisms through which ECHSs aim to increase high school graduation and college readiness and enrollment among students who are underrepresented in college, including first-generation, low-income and minority students.
First, they attempt to stimulate students' interest in higher education through exposure to college-style courses.
Second, they aim to build students' confidence about their ability to succeed in a postsecondary environment.
Third, they provide students opportunities to earn college credits in high school, thus lessening the future financial burden of college.
Additionally, ECHSs in North Carolina promote rigorous instructional practices, positive student-staff relationships and academic and student supports, which also may increase student achievement in and after high school.

Experimental research on ECHSs in North Carolina to date indicates that the ECHSs studied had a significant positive impact on a host of outcomes related to high school achievement and completion as well as college enrollment and degree attainment \citep{edmunds2016smoothing}. Of particular interest is success in ninth grade, given the compelling descriptive evidence that students who do well in ninth grade are more likely to succeed in and complete high school \citep{allensworth2005track}. \citet{edmunds2010preparing} found that being assigned to an ECHS had strong, positive impacts on ninth grade student outcomes.
Students given access to an ECHS were significantly more likely to enroll in and complete Algebra I, considered a gateway course for college-preparatory mathematics \citep{dougherty2015middle}, as well as have lower rates of school absence and suspension.
In addition, those offered the ECHS opportunity were five percentage points more likely to be ``on-track'' for college at the completion of ninth grade, meaning that they were on-track to complete the Future-Ready Core Graduation Requirements specified by the state of North Carolina.\footnote{The Future-Ready Core is designed to ensure that every student will graduate from high school prepared for the workforce and postsecondary education. For more information about the specific requirements, see: http://www.dpi.state.nc.us/docs/gradrequirements/resources/gradchecklists.pdf.}
These promising results beg for further investigation into where and for whom the ECHS model is particularly beneficial.

\subsection{Research question}
One possibility is that these top-line effects mask policy-relevant heterogeneity according to the type of school in which a student would enroll absent the ECHS offer. Indeed, the control students attended one of a number of different secondary schools that ranged substantially in overall quality, and we might believe that the variation in impact of the ECHS offer could be associated with variation across alternative contexts. Recent work in other domains has found similar relationships. Examining data from the Head Start Impact Study, \cite{FeGrMiPa2016} and \citet{kline2015evaluating} find that the impact of participating in Head Start, a large federally funded preschool program for low-income three and four-year-old children, varies according to the type of care a child would have otherwise received, with effects primarily for those who would be cared for in a home-based setting rather than those who would otherwise attend another child-care center. This is presumably because the shift from a home-based setting to the Head Start center setting represented a larger experiential change. Similarly, in a study of a school choice lottery in Charlotte-Mecklenburg, \citet{deming2014school} find that impacts of winning the lottery were largest for students living in areas with (and presumably who would otherwise attend) low-quality neighborhood schools.

In this spirit, we investigate variation in the impact on ninth-grade high school success of participating in an ECHS, according to variation in the quality of the high school in which ECHS participants would enroll absent the ECHS offer. This question is policy relevant. If, for example, the ECHS opportunity has a larger impact on outcomes for those who would otherwise attend a low-quality high school, then policymakers may prioritize locating new ECHSs in geographic settings where the existing public school opportunities are of low quality. Similarly, policymakers may see ECHSs as a viable reform strategy to improve faltering schools. Such variation also could motivate investigating quality of implementation of the ECHS model across different contexts. In short, our analytic goal is to differentiate treatment effects of ECHS participation for students who would otherwise attend a low-quality public school from those who would otherwise attend a high-quality public school.

\subsection{Data}
We utilize data from the Evaluation of Early College High Schools in North Carolina \citep{edmunds2010preparing}. Our sample includes 4,004 students across 6 cohorts who began ninth grade between 2005 and 2010.  Across these years, sample students participated in one of 44 lotteries to gain entry into one of 19 different ECHSs included in the ECHS evaluation. In each lottery, students randomized to treatment received an offer to attend an ECHS; those randomized to control did not receive an ECHS offer. Both treatment and control students were tracked through high school and college using data from the North Carolina Department of Instruction (NCDPI), the North Carolina Community College System and the National Student Clearinghouse.  We limited our sample to students who could be linked to NCDPI data, had school enrollment data in ninth grade and had transcript data or End or Course (EOC) exam data from NCDPI.
We also excluded the small number of students who we believe to have moved as the ninth grade school they attended was more than 20 miles away from the eighth grade school they attended, under the assumption that these large distances effectively dropped these students from the trial.
Our analysis sample under these business rules is 3,820 students.\footnote{Our overall attrition rate is 4.6 percent and differential attrition is 2.1 percent which puts us well within WWC attrition standards.  Including the predicted movers has little to no impact on the analyses presented in the paper.}

\subsection{School quality}
Up to this point, we have used the term ``school quality'' informally and recognize that many definitions are possible. The State of North Carolina uses a school report card system that considers a range of school-level measures, such as student performance on end-of-course tests, the share of students taught by ``high quality'' teachers, share of students taking the SAT, and various other growth indicators and scores.\footnote{See https://ssri.duke.edu/data-it-services/north-carolina-education-research-data-center-ncerdc.} The report card also includes a rating classification, one of nine categories based on achievement measures, growth and adequate yearly progress. For this study, we classify schools designated by the state as ``Priority Schools'', ``Low Performing Schools'' and ``Schools receiving no recognition'' as low-quality schools and classify those designated as ``Schools making High Growth,'' ``Schools making Expected Growth,'' ``Honor Schools of Excellence'', ``Schools of Excellence,'' and ``Schools of Progress'' as high-quality schools.\footnote{See http://www.ncpublicschools.org/docs/accountability/reporting/abc/2005-06/execsumm.html for detailed descriptions of categories. The state does not rate special education schools, schools in hospitals, vocational schools and schools with unresolved data. We grouped all schools without a rating into their own category. Among schools we classify as high-quality, the majority (83\%) are classified by the state as ``Schools of Distinction,'' and among those that we classify as low-quality, the majority (89\%) are rated by the state as ``Schools of Progress.'' } With designations collapsed in this way, we are able to categorize students as having attended an ECHS, a low-quality school, or a high-quality school.\footnote{ It is worth noting that ECHSs also receive quality ratings from the state, and some ECHS schools in our sample are rated as low-quality schools. However, we treat ECHSs as their own quality category because an ECHS is guided by different principles than a traditional school and provides students with a different experience that may not be captured in school ratings.}

In Table~\ref{tab:schltype_dist}, we present the distribution of students across these three school settings in their ninth grade year. In the treatment group, 86 percent of students attended an ECHS, 2.5 percent attended a high-quality school, and 11.5 percent a low-quality school.  In the control group, only 3 percent were able to cross over and register in an ECHS; 14 percent attended a high-quality school; and 83 percent attended a low-quality school. Next, we turn to the specific principal stratification set up that frames our investigation.\\

\begin{table}[hbt]
\center
\begin{tabular}{lcc}
School type	& Treatment & Control \\
    & (N = 2,229) & (N = 1,591) \\
\hline
Early College HS ($e$) &	86.0\% &	2.9\% \\
High Quality Public HS ($hq$) &	2.5\%	& 13.9\% \\
Low Quality Public HS ($lq$) &	11.5\%	& 83.2\% \\
\hline
\end{tabular}
\caption{Distribution of high school type attended by treatment status}
\label{tab:schltype_dist}
\end{table}

\section{Defining Principal Strata}

To define our principal strata we first introduce notation.
We use a finite population framework with potential outcomes (\cite{SplawaNeyman:1990ux}, see also, e.g., \cite{imbens2015causal}).
In this framework we take as our entire population the $N$ students in our sample, indexed with $i = 1, \ldots, N$, on which the experiment was conducted.
We take the finite sample approach to explicitly focus on the primary source of randomness---the treatment assignment itself---and to avoid the need to make assumptions of the sample being representative of some larger infinite population.
In our case it also makes the definition of the estimands of interest more explicit, especially when we stratify by baseline covariate.
For further general discussion see, for example, \cite{Rosenbaum:2012ul} or \cite{imbens2015causal}.
To continue, let $Y_i(0), Y_i(1)$ be student $i$'s \emph{potential outcomes}, given possible treatment assignments $Z_i=0$ or $Z_i=1$.
We assume the Stable Unit Treatment Value Assumption (SUTVA), which states that the observed outcomes of a student are fully described by that student's assignment and not the assignment of the student's peers \citep{rubin1980randomization}.\footnote{SUTVA here, as in most non-compliance settings, is a bit subtle in that it is regarding treatment offer. The additional formalism around treatment take-up as potential outcomes in their own right allow us to identify the impact of treatment receipt (school attendance). We also note that the non-interference is somewhat suspect due to students being peers in different schools; we follow much of the education literature in assuming these interference impacts are minimal when looking at subsets of students in larger institutions, as we are doing.}
Similarly, what school type a child attends also depends on treatment, so we consider this post-treatment behavior a potential outcome in its own right.
In particular, in the context of this study, school type attended can take on one of three levels: ECHS ($e$), low-quality ($lq$), or high-quality ($hq$).
Then let $S_i(z) \in \{e, lq, hq\}$ be the categorical variable of what school type student $i$ attends if assigned treatment $Z_i=z$.

These $S_i(z)$ variables are key: we define our principal strata by the joint distribution of these intermediate variables under treatment and control.
In the ECHS context, for example, the principal strata are defined by the pair of potential school types students would attend if assigned to treatment, $S_i(1)$, and if assigned to control, $S_i(0)$.

Given our three different focal schools types, there are at most nine different strata, represented by the nine cells in Table~\ref{tab:princ_strat}.
Each student belongs to exactly one of these cells, depending on that student's $S_i(0)$ and $S_i(1)$.
For example, a student who would go to an ECHS school under treatment ($S_i(1)=e$), and a low quality school under control ($S_i(0) = lq$), would be in stratum (2).
The strata contain distinct groups of students, and each of these groups has its own stratum-specific average treatment impact.
To simplify the problem, we next apply two assumptions that serve to reduce the number of possible strata from nine to five.

The first assumption is the monotonicity assumption, which states that there are no students who would enroll in an ECHS when not offered access but who would not take up the opportunity to enroll when offered.

\begin{myassump}{1}
\label{assump:monotonicity}\textit{(Monotonicity/No Defiers.)} No individuals have $S_{i}(0) = e$ and $S_{i}(1) \ne e$.
\end{myassump}
Due to this assumption we rule out the existence of strata 4 and 7 (see Table~\ref{tab:princ_strat}).

The second assumption states that the ECHS offer does not change the school setting for students choosing between non-ECHS options. For example, the offer does not induce students to switch from a low-quality to a high-quality school.  

\begin{myassump}{2}\textit{(Irrelevant Alternatives.)}\label{assump:irrel_alt} No individuals have $S_i(0) = hq$ and $S_i(1) = lq$ or $S_i(0) = lq$ and $S_i(1) = hq$.
\end{myassump}

The underlying reasoning here is that since the availability of other schooling options is not affected by the ECHS offer, preferences among the non-ECHS options should not be affected by an offer to attend an ECHS school. Thus, we rule out strata 6 and 8.

We now have five remaining principal strata: ECHS always-takers (EAT), low-quality compliers (LQC), high-quality compliers (HQC), low-quality always-takers (LQAT), and high-quality always-takers (HQAT), illustrated in Table~\ref{tab:princ_strat}.
We encode individual strata membership with, for each student, $R_i \in \{ eat, lqat, lqc, hqat, hqc \}$.
$R_i$ is a function of the pair $(S_i(0), S_i(1))$.

\begin{table}[hbt]
\center
\begin{tabular}{llr|c|c|c}
	&		&  			& \multicolumn{3}{c}{NO ECHS offer ($Z_i=0$)} \\
	&		&			& $S_i(0)=e$ 		& $S_i(0)=lq$  			& $S_i(0)=hq$  \\
  \cline{1-6}
 \parbox[t]{2mm}{\multirow{3}{*}{\rotatebox[origin=c]{90}{ECHS offer\;\;}}  }  &
 \parbox[t]{2mm}{\multirow{3}{*}{\rotatebox[origin=c]{90}{($Z_i=1$)\;\;\;\;}}}    &	$S_i(1)=e$& (1) ECHS  	& (2) Low-quality  	& (3) High-quality  \\
  	&	 	&			&  	always taker   &  complier 	    &  complier \\
 \cline{3-6}
  	&		 &$S_i(1)=lq$	& 	(4)		    &(5) Low-quality   	    & (6) \\
 	&		 &			& 			    & always taker	    & \\
 \cline{3-6}
 	&		 &$S_i(1)=hq$	&	(7)		    &		(8)		    & (9)  High-quality \\
	&		 &			&			    &				    & always taker \\
	&		 & 			&				&					&
\end{tabular}
  \caption[Diagram of principal strata]{Our principal strata. We have nine possible strata given the three possible school-types a student could attend under treatment and control. We assume that no students fall in strata (4), (6), (7) and (8), as discussed in the text, leaving us with the five remaining groups.}
  \label{tab:princ_strat}
\end{table}

With our principal strata defined, we can turn to defining our parameters of interest.
In particular, we are interested in the share of students in each stratum, $\pi_r$, and the stratum outcome means under treatment and control, $\mu_{r}(z)$.
The relative strata proportions are $\pi_r = N_{r}/N$, where $N_{r}$ is the number of students in stratum $r$.
The mean outcomes under both treatment and control are
\[ \mu_{r}(z) = \frac{1}{N_r} \sum_{i=1}^N \ind{R_i = r} Y_i( z ) , \]
where $\ind{ A = a }$ is the indicator function evaluating to 1 if $A = a$ and 0 otherwise.
All these quantities are fixed, pre-treatment.

In total, there are $5 \times 3 = 15$ population parameters of interest (a $\pi_r$, $\mu_r(1)$, and $\mu_r(0)$ for each stratum), and our primary substantive goal is to estimate treatment effects of ECHS participation for low-quality and high-quality compliers.
This effect for the LQC group is
\begin{align*}
 ITT_{lqc} &= \mu_{lqc}(1) - \mu_{lqc}(0) \\
    &= \frac{1}{ N_{lqc} } \sum_{i=1}^N \ind{ R_i = lqc } \left( Y_i(1) - Y_i(0) \right) .
\end{align*}
$ITT_{hqc}$ is analogous.

\subsection{Defining our directly estimable quantities}
The above determines the parameters, but we of course do not observe these quantities directly.
Instead, after individuals are randomized to treatment or control, for each student $i$ we observe three variables: $Z_i$, an indicator of treatment; $S_i^{obs} = S_i(Z_i)$, the observed school type attended; and $Y_i^{obs} = Y_i(Z_i)$, student $i$'s observed outcome. 
We also know each individual's probability of treatment assignment, which is captured in a weight $w_i$ (we return to a discussion of weights later in the paper).
Unfortunately, we do \emph{not} fully observe $R_i$, individual stratum membership, because that would require seeing both $S_i(0)$ and $S_i(1)$, and we only see one of these.
This means we only partially know which stratum any particular student is in.
We do, however, observe groups defined by  $Z_i$ and $S_i^{obs}$, which are mixtures of students from different strata.
Therefore, outcome distributions within these observed groups are mixtures of outcomes across different principal strata. The goal of estimation and bounding is, in essence, to separate out these mixtures.

In particular, we have six groups based on the combination of treatment assignment and subsequent school attended.
We index these groups with $zs$, with $z \in \{0, 1\}$ and $s \in \{ e, lq, hq \}$.
Each student will be a member of \emph{two} of these groups, one for each potential treatment, and membership in one group will be directly observable for each student.
Due to random assignment, we obtain random samples of all six groups, meaning we can estimate the characteristics of these groups.
In other words, the parameters associated with these groups are directly identifiable, and thus we use them as the building blocks for estimating the actual parameters of interest pertaining to the five principal strata.

First, we can directly estimate the number of students who, if assigned treatment $z$, go to school-type $s$:
\[  N_{zs} = \sum_{i=1}^N \ind{ S_i(z) = s }. \]

We then have $\rho_{zs} = N_{zs} / N$,  the proportion of all the students who would go to school-type $s$ if assigned treatment $z$.\footnote{We divide by $N$ because these $N_{zs}$ are the share of the \emph{entire} sample that would, if the entire sample were assigned treatment $z$, end up at school type $s$. Under a finite sample model, each student will be in two of the groups defined by $z$ and $s$, and $N_{zlq} + N_{zhq} + N_{ze} = N$ for both $z = 0$ and $z = 1$.} 
For example, $\rho_{1e}$ is the proportion of the students who, when offered the ECHS opportunity, actually enroll in an ECHS school.
For clarity note:
\[ \rho_{ze} + \rho_{zlq} + \rho_{zhq} = 1 \mbox{ for } z = 0, 1 . \]

Similarly, let $\overline{Y}_{zs}$ be the mean outcomes of all those who, if assigned to treatment $z$, would go to school-type $s$:
\[ \overline{Y}_{zs} = \frac{1}{N_{zs}} \sum_{i=1}^N \ind{ S_i(z) = s } Y_i(z) . \]

The $\rho_{zs}$ and $\overline{Y}_{zs}$ are parameters of our finite population.
Because of random assignment and SUTVA, we can estimate these quantities directly.
Throughout the subsequent discussion on identifiability we therefore assume that we have full knowledge of these intermediate $\rho_{zs}$ and $\overline{Y}_{zs}$.
These are our building blocks, and we characterize the $\pi_{r}$, the $\mu_{r}(z)$, and finally the $ITT_{r}$ in terms of these values.
We purposefully defer issues of estimation and uncertainty to Section~\ref{sec:estimation}.
We next discuss calculating the $\pi_{r}$, and then the $\mu_{r}(z)$.

\subsection{Identifying the proportions}
In this principal stratification set up, we are able to point identify the distribution of students across the five strata of interest.
Assuming we can identify the $\rho_{zs}$, we are able to identify $\pi_r$ for $r \in \{ eat, lqat, lqc, hqat, hqc \}$.
More generally, if strata of interest are defined by a cross-tabulation of $x$ rows and $y$ columns, then several parameters, such as the share of student in each stratum and the average value of baseline characteristics, are point identified if $(x-1) \times (y-1)$ cells are eliminated from possibility and at least one cell remains in each row and in each column.
In Table~\ref{tab:princ_strat} these conditions are met.

In the ECHS experiment, students who are offered ECHS enrollment either accept or not. If they do not, we observe the school type that they select instead.
This allows direct estimation of the proportion of low- and high-quality always-takers ($\pi_{lqat} = \rho_{1lq}$ and $\pi_{hqat} = \rho_{1hq}$).

Students, when not offered treatment, either cross over into an ECHS or stay in their respective quality schools.
This allows direct estimation of the proportion of ECHS always-takers as $\pi_{eat} = \rho_{0e}$.
Those students who stay in their respective school types under control assignment are combinations of two different types of students.
For example, those students who would enroll in a low-quality public school under control are either low-quality compliers or low-quality always-takers.
This leads to the following equalities:
\[ \rho_{0lq} = \pi_{lqc} + \pi_{lqat} \]
and
\[ \rho_{0hq} = \pi_{hqc} + \pi_{hqat} . \]
We then plug in our known quantities to obtain the other.

In sum, we can obtain $\pi_{lqat}$, $\pi_{hqat}$, and $\pi_{eat}$ directly and can express the remaining stratum proportions as functions of observable values, as follows:
\begin{align*}
 \pi_{lqc} &= \rho_{0lq} - \pi_{lqat} = \rho_{0lq} - \rho_{1lq} \mbox{ and } \\
 \pi_{hqc} &= \rho_{0hq} - \pi_{hqat} = \rho_{0hq} - \rho_{1hq} .
\end{align*}

\subsection{Identifying individual stratum outcome means}
We obtain stratum-specific treatment effects by obtaining the treatment and control means, $\mu_{r}(z)$, which make up the treatment contrasts.
To do so, we first apply an additional assumption that serves to constrain certain means.
This allows us to disentangle observed mixtures of different student types.

\begin{myassump}{3}\textit{(Exclusion restrictions.)}\label{assump:exlusion} For students with $R_i \in \{eat, lqat, hqat\}, Y_i(0) = Y_i(1)$.
\end{myassump}

This assumption states that there is no impact of the randomized offer of ECHS enrollment for students whose school type is unchanged by randomization.
This implies that the treatment effect is zero in these three strata.
Furthermore, this means the associated stratum-specific means are point identified because each type of always-taker is isolated in one or the other of the treatment arms.
Therefore we have
\begin{align*}
 \mu_{lqat}(0) &= \mu_{lqat}(1) = \mu_{lqat} = \overline{Y}_{1lq} \\
 \mu_{hqat}(0) &= \mu_{hqat}(1) = \mu_{hqat} = \overline{Y}_{1hq}\\
 \mu_{eat}(0) &= \mu_{eat}(1) = \mu_{eat} = \overline{Y}_{0e} .
\end{align*}

Next, because the group of students who experience a low-quality school setting under assignment to control is a mixture of low-quality compliers and low-quality always-takers, the mean outcome of these students is a weighted average of the respective mean control outcomes of the two principal strata, weighted by their relative sizes:
\[ \overline{Y}_{0lq} = \frac{ \pi_{lqc} }{ \pi_{lqc} + \pi_{lqat} } \mu_{lqc}(0) + \frac{ \pi_{lqat} }{ \pi_{lqc} + \pi_{lqat} } \mu_{lqat}(0) . \]
We rearrange to get
\[  \mu_{lqc}(0) = \frac{ \pi_{lqc} + \pi_{lqat} }{ \pi_{lqc} } \overline{Y}_{0lq} -  \frac{ \pi_{lqat} }{ \pi_{lqc} } \mu_{lqat}(0).  \]

Plugging in $\mu_{lqat}(0) = \overline{Y}_{1lq}$ and our proportion relations, we can express our control mean as a combination of directly estimable quantities:
\begin{equation}
 \mu_{lqc}(0) = \frac{\rho_{0lq}}{\rho_{0lq} - \rho_{1lq}} \overline{Y}_{0lq} - \frac{\rho_{1lq}}{\rho_{0lq} - \rho_{1lq}}\overline{Y}_{1lq} . \label{eq:control-mean}
\end{equation}

The expression for $\mu_{hqc}(0)$ is analogous, with the $\rho_{zlq}$ and $\overline{Y}_{zlq}$ replaced by the $\rho_{zhq}$ and $\overline{Y}_{zhq}$.

We have directly obtained the mean control outcome for our two groups of interest.
We can also obtain the average of our two groups of interest by averaging the above two means weighted by the relative sizes of the two groups:
\[ M_0 = \frac{\pi_{lqc}}{\pi_{lqc}+\pi_{hqc}} \mu_{lqc}(0) + \frac{\pi_{hqc}}{\pi_{lqc}+\pi_{hqc}} \mu_{hqc}(0) . \]
This would be the \emph{overall} complier mean outcome under control.

\paragraph{Means of covariates within strata.}
We can also identify the mean of any pre-treatment covariate $X_i$ for all five strata using the various formula for the $\mu_{r}(0)$ by replacing the observed means $\overline{Y}_{zs}$ with analogous $\overline{X}_{zs}$.
In particular, we first estimate the means of the groups we directly identify, and then back out the mean of the two complier groups using Equation~\ref{eq:control-mean}.

\paragraph{The treatment means of the compliers.}
The only outcome means that remain are the means for low-quality and high-quality compliers under assignment to treatment ($\mu_{lqc}(1)$ and $\mu_{hqc}(1)$). Students who take up the ECHS offer under assignment to the treatment are a mixture of three types of students, and so the observed mean is a weighted average of these groups:
\begin{equation}
 \overline{Y}_{1e} = \frac{ \pi_{eat} }{\pi_{eat}+\pi_{lqc}+\pi_{hqc} } \mu_{eat} + \frac{ \pi_{lqc} }{\pi_{eat}+\pi_{lqc}+\pi_{hqc} } \mu_{lqc}(1) + \frac{ \pi_{hqc} }{\pi_{eat}+\pi_{lqc}+\pi_{hqc} } \mu_{hqc}(1) . \label{eq:mean-1e}	
\end{equation}
The exclusion restriction allows us to obtain $\mu_{eat} = \overline{Y}_{0e}$, but we cannot separate the two means of interest, $\mu_{lqc}(1)$ and $\mu_{hqc}(1)$. These parameters are those that are not point identified but are set identified.
That is, we will be able to obtain upper and lower bounds for these parameters and also characterize the trade-off between the two, but we cannot obtain them exactly.

We can, however, identify the \emph{overall} complier mean outcome under treatment, $M_1$, which is simply a weighted average between $\mu_{lqc}(1)$ and $\mu_{hqc}(1)$:
\begin{align}
 M_1 &\equiv \frac{ \pi_{lqc} }{ \pi_{lqc}+\pi_{hqc}} \mu_{lqc}(1) + \frac{ \pi_{hqc} }{ \pi_{lqc}+\pi_{hqc}} \mu_{hqc}(1) \label{eq:M1}
\end{align}
By rearranging Equation~\ref{eq:mean-1e} by moving the term with $\mu_{eat}$ to the left side and then rescaling both sides by $(\pi_{eat}+\pi_{lqc}+\pi_{hqc})/(\pi_{lqc}+\pi_{hqc})$ we have
\begin{align*}
 M_1  &= \frac{\pi_{eat}+\pi_{lqc}+\pi_{hqc}}{ \pi_{lqc}+\pi_{hqc}} \overline{Y}_{1e} - \frac{ \pi_{eat} }{ \pi_{lqc}+\pi_{hqc}} \mu_{eat} \\\\
   &=  \frac{ \rho_{1e}} { \rho_{1e} - \rho_{0e} } \overline{Y}_{1e} - \frac{ \rho_{0e} }{ \rho_{1e} - \rho_{0e} } \overline{Y}_{0e} .
\end{align*}
The last step comes from plugging in earlier expressions for the $\pi_r$ and $\mu_{eat}$ terms.

$M_1$ is a weighted average of our two means of interest, which allows us to characterize possible trade-offs.
In particular, if one stratum mean is large, the other must be small.
Overall, to bound $\mu_{lqc}(1)$ and $\mu_{hqc}(1)$ we explore possible combinations with a weighted average equal to this $M_1$.

\subsection{Illustration on ECHS Data}
To make this process concrete, we turn to the ECHS data (which we discuss more fully in Section~\ref{sec:application}).
Our outcome of interest is a binary indicator for whether a student is on-track at the end of ninth grade.
We first calculate the share of students within each observable group defined by treatment assignment and type of school attended.
Within these groups, we calculate the average value of the outcome.
We present these results in Table~\ref{tab:observed_data}.

These quantities are the actual number of students, estimated mean outcomes, and estimated overall proportions for each group.
Students were weighted according to their differential probabilities of receiving treatment, which sometimes differed by demographic characteristics \citep{edmunds2016smoothing}, and the estimates incorporate these weights (we discuss how to do this in Section~\ref{sec:estimation}).
The random assignment mechanism means the students observed to be in a given group are a random sample of all students who could be in the group, which allows for this estimation.

\begin{table}[ht]
\centering
\begin{tabular}{rrlrrr}
Group & $Z$ & $S$ & $N_{zs}$  & $\widehat{Y}_{zs}$ & $\hat{p}_{zs}$ \\
  \hline
$0e$   &   0   &   e   &   43      &   100   &  3\\
$0hq$   &   0   &   hq   &   220     &   97   &   14\\
$0lq$   &   0   &   lq   &   1328     &   86   &   83\\
$1e$   &   1   &   e   &   1917     &   95   &   86\\
$1hq$   &   1   &   hq   &   53     &   90   &   3\\
$1lq$   &   1   &   lq   &   259     &   84   &   12\\
   \hline
\end{tabular}
\caption[Observed subgroup outcomes for ECHS study]{Observed subgroup outcomes for ECHS study. $Z$ and $S$ divide our population into 6 groups. The $N_{zs}$ are the total number of students in each group. $\widehat{Y}_{zs}$ (in percents) are the estimated mean outcomes of each group (i.e., the percent of the group on-track). They estimate the $\overline{Y}_{zs}$.  The $\hat{p}_{zs}$ (in percents), estimating $\rho_{zs}$, are the estimated fractions of the groups in the given treatment arms.}
\label{tab:observed_data}
\end{table}

Our analytic goal is to estimate the 15 stratum-specific parameters, $\pi_r$, $\mu_{r}(0)$, and $\mu_{r}(1)$ for $r$ $\in$ $\{eat, hqat, lqat, hqc, lqc\}$.
As articulated above, the sample results in Table~\ref{tab:observed_data} allow us to estimate 13 of the 15 parameters.
These estimates are in the left three columns of Table~\ref{tab:identified_estimates}.
We cannot, however, directly estimate the remaining two parameters, $\mu_{lqc}(1)$ and $\mu_{hqc}(1)$.

One thing we can do is estimate the overall complier outcome mean under treatment and control.
We estimate $\widehat{M}_1=0.95$ and $\widehat{M}_0 = 0.88$, for an overall estimated complier average causal effect of $0.07$, or a 7 percentage point change in the on-track status of the students.
Analytically, our goal is to disentangle this overall complier average causal effect into separate effects for those compliers who would otherwise attend low-quality schools and those who would otherwise attend high-quality schools.
Since the treatment group means contributing to these separate effects are not point identified, we bound them based on directly observed quantities.

The fourth and final columns show the bounds we obtain for these quantities, and the associated bounds on the ITT estimates themselves.
Although these bounds are fairly wide, they are nevertheless informative. 
The treatment impacts on the compliers who would attend a low-quality high school absent the ECHS offer are definitively positive and substantial, indicating that ECHS participation increases on-track status by somewhere between 8 to 14 percentage points for these students.
The effects for the high-quality compliers are less encouraging. 
Although these may be slightly positive, they could also be quite negative, potentially reducing the share of students on track by 34 percentage points. 

Figure~\ref{fig:trade-off-figure}(b) illustrates the potential tradeoffs possible: if the HQC effect is substantially negative, the LQC effect is relatively larger, and vice-versa.  
All possible pairs of effects must lie on the line segment corresponding to an overall complier average effect of $M_1 - M_0$.
We discuss how to obtain these bounds next.

\begin{table}[ht]
\centering
\begin{tabular}{lrrc|c}
  Stratum & $\pi_r$ & $\mu_{r}(0)$ & $\mu_{r}(1)$ & $ITT_{r}$  \\
  \hline
eat & 0.03 & 1.00 & 1.00 &   0   \\
   hqat & 0.03 & 0.90 & 0.90 & 0 \\ 
   hqc & 0.11 & 0.99 &  $(0.64, 1.00)$ &  $(-0.34, 0.01)$  \\
   lqat & 0.12 & 0.84 & 0.84 & 0 \\ 
   lqc & 0.72 & 0.86 & $(0.94, 1.00)$ & $(0.08, 0.14)$  \\
   \hline
\end{tabular}
\caption[Estimated population parameters for ECHS study]{Estimated population parameters for ECHS study with corresponding average treatment effects.  Ranges (in parentheses) are from bounding parameters that are not point-identified.}
\label{tab:identified_estimates}

\end{table}

\section{Calculating bounds in terms of directly observed quantities}
We know the weighted average of $\mu_{lqc}(1)$ and $\mu_{hqc}(1)$ equals our observed overall complier mean, as shown on Equation~\ref{eq:M1}.
We also know that each mean must lie in the $[0,1]$ interval because our outcome of interest is binary, i.e., $Y_i \in \{ 0, 1 \}$.
The restricted range of $Y_i$ means the $\mu_r(1)$ cannot be dominated by extreme values, making this bounding exercise possible.
Bounds with continuous $Y_i$ are possible in some circumstances; see, for example, \cite{Lee:2009tc}.

\begin{figure}[hbt]
\centering
\begin{subfigure}[b]{.45\textwidth}
  \centering
  \includegraphics[width=.9\linewidth]{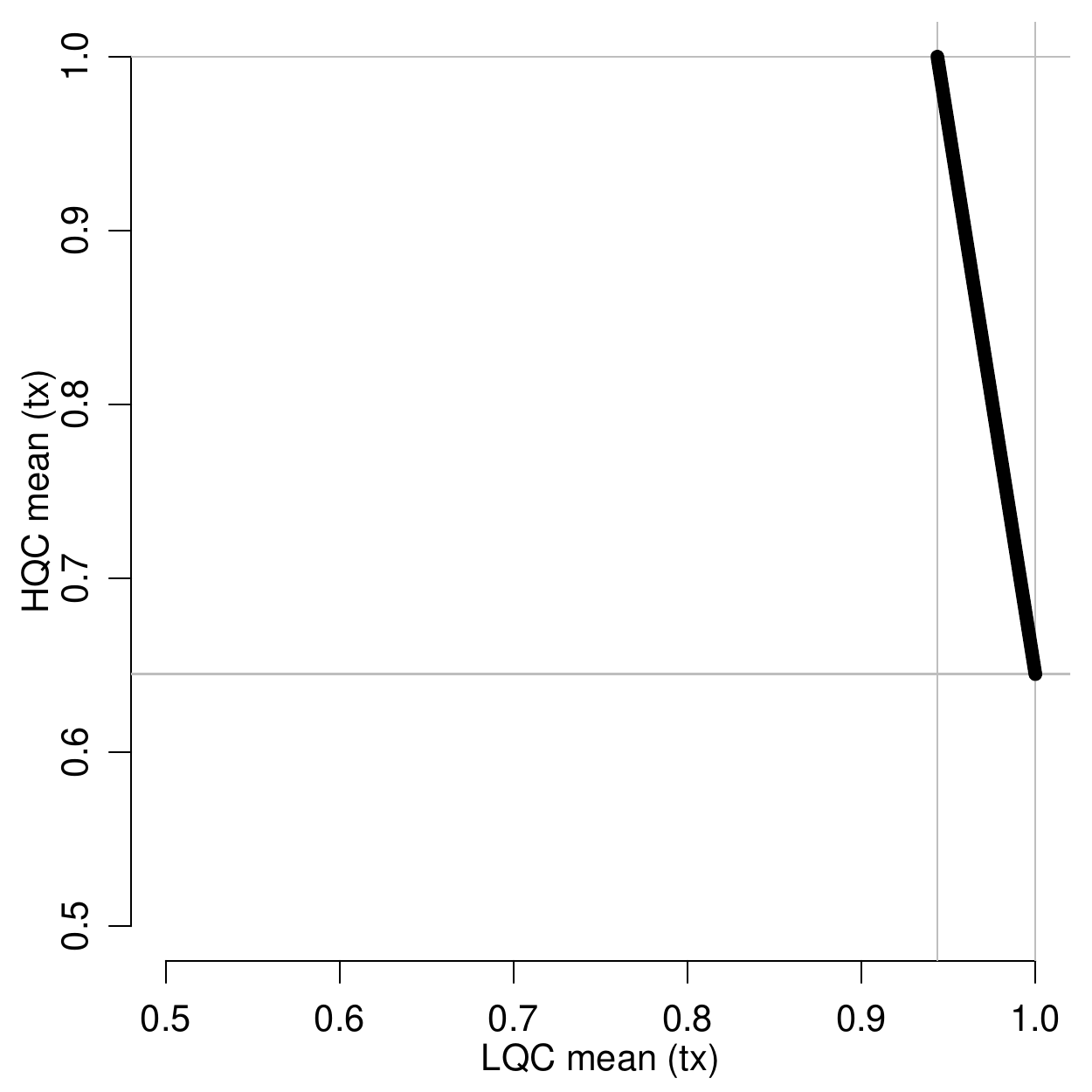}
  \caption{Treatment means}
\end{subfigure}
\begin{subfigure}[b]{.45\textwidth}
  \centering
  \includegraphics[width=.9\linewidth]{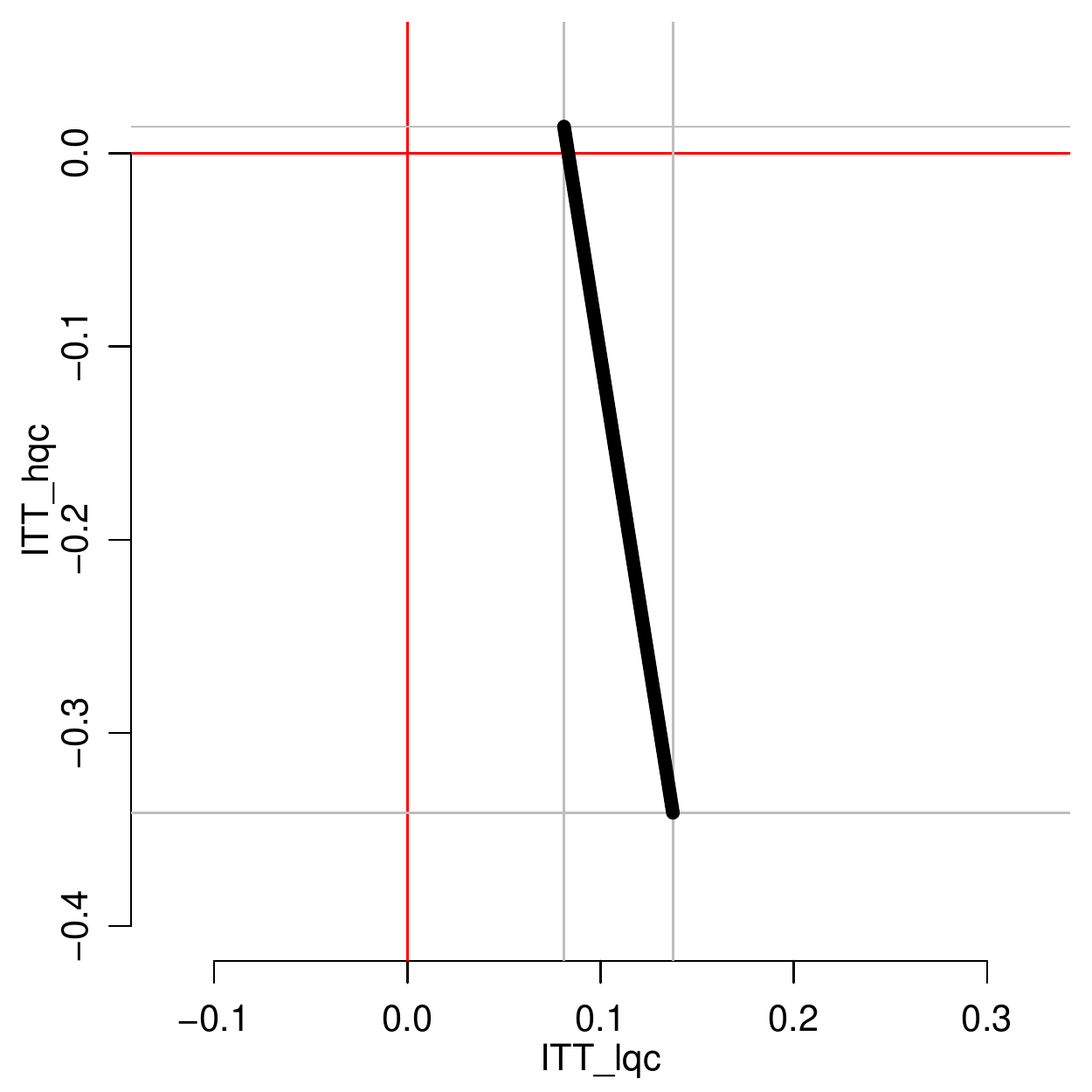}
  \captionof{figure}{Treatment effects}
\end{subfigure}
\caption[Possible trade-offs between the treatment means and treatment effects]{The possible trade-offs between the treatment means and treatment effects of the LQC and HQC group.  On left, each point on the dark segment represents a possible value for the pair $\mu_{lqc}(1), \mu_{hqc}(1)$ that would give an overall mean of $M_1 = 0.95$.  The right hand side is simply the left shifted by the control means of each group.}
\label{fig:trade-off-figure}
\end{figure}

To get expressions of the bounds we solve for one of our means of interest.
For example, when we solve Equation~\ref{eq:M1} for $\mu_{lqc}(1)$ we get
\begin{align}
  \mu_{lqc}(1)  &= \frac{\pi_{lqc}+\pi_{hqc}}{\pi_{lqc}}  M_1 - \frac{\pi_{hqc}}{\pi_{lqc}} \mu_{hqc}(1) \label{eq:lqc-equation}
\end{align}
This shows, for example, that for $\mu_{lqc}(1)$ to be non-negative, $\mu_{hqc}(1)$ must be no larger than $\frac{\pi_{lqc}+\pi_{hqc}}{\pi_{hqc}}  M_1$.
We can further explore the range of possible trade-offs by plugging different possible values of $\mu_{hqc}(1)$ into Equation~\ref{eq:lqc-equation}.
This is shown on Figure~\ref{fig:trade-off-figure}(a).
The overall bounds come from considering what happens to each mean when the other mean is maximal or minimal, coupled with restricting each mean to the $[0,1]$ interval.

For the low-quality compliers, we have bounds of
\begin{align*}
 \max \left( 0, \frac{ \pi_{lqc} + \pi_{hqc} }{ \pi_{lqc} }M_1 - \frac{\pi_{hqc}}{\pi_{lqc}} \right) \leq \mu_{lqc}(1) &\leq  \min \left( \frac{ \pi_{lqc} + \pi_{hqc} }{ \pi_{lqc} }M_1, 1 \right) .
\end{align*}
The derivation of the lower bound corresponds to plugging $\mu_{hqc}(1)=1$ into Equation~\ref{eq:lqc-equation}: the lowest possible bound for the LQC group is when the mean for the HQC group is maximal.

For the high-quality compliers, by symmetry, we can swap the $lqc$ with the $hqc$ in the above expressions.

\paragraph{The final bounds.}
With bounds on these stratum means under treatment, we are able to obtain bounds on the stratum-specific treatment effects for these two groups.
For $ITT_{lqc} = \mu_{lqc}(1) - \mu_{lqc}(0)$ we have
\begin{equation}
 \mu_{lqc}(1)^{low} - \mu_{lqc}(0) \leq ITT_{lqc} \leq \mu_{lqc}(1)^{high} - \mu_{lqc}(0) \label{eq:bounds}	
\end{equation}
with
\begin{align*}
\mu_{lqc}(1)^{low} &\equiv \max \left( 0, \frac{ \pi_{lqc} + \pi_{hqc} }{ \pi_{lqc} }M_1 - \frac{\pi_{hqc}}{\pi_{lqc}} \right)  \\
\mu_{lqc}(1)^{high} &\equiv \min \left( \frac{ \pi_{lqc} + \pi_{hqc} }{ \pi_{lqc} }M_1, 1 \right) \\
\intertext{and}
\mu_{lqc}(0) &\equiv \frac{ \pi_{lqc} + \pi_{lqat} }{ \pi_{lqc} } \overline{Y}_{0lq} -  \frac{ \pi_{lqat} }{ \pi_{lqc} } \mu_{lqat} .
\end{align*}
The bounds are a rescaled $M_1$ separated by $\pi_{hqc}/\pi_{lqc}$ and cropped by 0 and 1.

To get these in our directly identifiable quantities, plug in our expressions for the $\rho_{zs}$ and remaining $\overline{Y}_{zs}$ to get
\begin{align*}
\mu_{lqc}(1)^{low} &= \max \left( 0, \frac{ \rho_{1e} - \rho_{0e} }{ \rho_{0lq} - \rho_{1lq} } M_1 - \frac{\rho_{0hq}-\rho_{1hq}}{\rho_{0lq}-\rho_{1lq}} \right)  \\
\mu_{lqc}(1)^{high} &= \min \left( \frac{ \rho_{1e} - \rho_{0e} }{ \rho_{0lq} - \rho_{1lq} } M_1, 1 \right)  \\
\mu_{lqc}(0) &= \frac{ \rho_{0lq} }{ \rho_{0lq} - \rho_{1lq} } \overline{Y}_{0lq} -  \frac{ \rho_{1lq} }{ \rho_{0lq} - \rho_{1lq} } \overline{Y}_{1lq} \\
\intertext{with}
M_1 &=\frac{ \rho_{1e}} { \rho_{1e} - \rho_{0e} } \overline{Y}_{1e} - \frac{ \rho_{0e} }{ \rho_{1e} - \rho_{0e} } \overline{Y}_{0e}  .
\end{align*}

The width of the bound is governed by $\pi_{hqc}/\pi_{lqc} = (\rho_{0hq}-\rho_{1hq})/(\rho_{0lq}-\rho_{1lq})$, the ratio of high- to low-quality compliers.
This makes sense: if the overall group is mostly low-quality compliers, resulting in a small ratio, then the LQC average effect has to be close to the overall average effect.
It also shows that if $M_1$ is close to 1, the inflation factor will scale $M_1$ above 1, so it gets cropped, giving a shorter bound.
Similarly, if $M_1$ is close to 0, then the rescaled $M_1$ minus the $\pi_{hqc}/\pi_{lqc}$ term will be negative, and the bound will therefore get cropped below by 0, also shortening it.

For the high quality compliers we analogously have $ITT_{hqc} = \mu_{hqc}(1) - \mu_{hqc}(0)$ with
\begin{align*}
\mu_{hqc}(1)^{low} &= \max \left( 0, \frac{ \rho_{1e} - \rho_{0e} }{ \rho_{0hq} - \rho_{1hq} } M_1 - \frac{\rho_{0lq}-\rho_{1lq}}{\rho_{0hq}-\rho_{1hq}} \right)  \\
\mu_{hqc}(1)^{high} &= \min \left( \frac{ \rho_{1e} - \rho_{0e} }{ \rho_{0hq} - \rho_{1hq} } M_1, 1 \right)  \\
\mu_{lqc}(0) &= \frac{ \rho_{0hq} }{ \rho_{0hq} - \rho_{1hq} } \overline{Y}_{0hq} -  \frac{ \rho_{1hq} }{ \rho_{0hq} - \rho_{1hq} } \overline{Y}_{1hq} .
\end{align*}
For the high quality compliers, the width is governed by $\pi_{lqc}/\pi_{hqc}$.

These bounds are sharp in that they are the shortest possible given the directly identified quantities.
In particular, given these bounds one could assign strata memberships so as to be entirely consistent with either extreme.
Conceptually this can be done by assigning those individuals with better outcomes to one of the two groups until that group is ``full,'' 
and then calculating the outcome for the other group with the individuals that remain.
Furthermore, these overall bounds' widths are dominated by the ratio of the proportions of high to low quality compliers.
In particular, if a group represents a relatively small proportion of the compliers, it will tend to have very wide bounds.
Bound width is \emph{not} related to sample size.
Sample size would impact uncertainty, however, and in small samples this uncertainty could be of greater magnitude than the bound widths (see Section~\ref{sec:estimation}).

We address the estimation of uncertainty in all of these estimates in Section~\ref{sec:estimation}.
With the goal of restricting the range of possible values for the groups we now turn to narrowing bounds using predictive covariates.

\subsection{Narrowing bounds through covariate stratification}
We tighten our bounds by exploiting pre-treatment covariates predictive of outcome and/or post-treatment behavior.
In particular, we will slice our overall sample of $N$ students into $K$ \emph{slices}, with these slices being defined by pre-treatment covariates and thus being conceptually fixed prior to randomization.
This type of analytic process is typically referred to as covariate (or post) stratification.
For clearer prose regarding principal stratification and covariate stratification, here we say \emph{slice} rather than \emph{stratum} to refer to the subsets of the sample defined by observable baseline covariate values.  Each slice is then its own sub-experiment on $N_k$ students, which we can assess just as with a single overall experiment. We will calculate bounds within these separate slices of the data, and then aggregate across them to get the overall bound. As we show, this can be an effective strategy for tightening bounds, provided the covariates are sufficiently predictive of post-treatment behavior and/or the outcome.

We discuss estimating effects for the LQC group.
The procedure for the HQC group is analogous.
First, we represent slice membership with a pre-treatment membership variable $X_i \in \{ 1, \ldots, K\}$, which we observe. Within each slice $k$, we wish to obtain the LQC average treatment effect:
\[  ITT_{lqc}^k = \frac{1}{N_{lqc,k} } \sum_{i: X_i = k} \ind{ R_i = lqc } \left( Y_i(1) - Y_i(0) \right), \]
with $N_{lqc,k} = N_k \pi_{lqc}^k$, the total number of LQC students in slice $k$.

The overall $ITT_{lqc}$ across slices can then be expressed as
\begin{align}
ITT_{lqc} &= \frac{1}{N_{lqc}} \sum_{k=1}^K N_{lqc,k} \frac{1}{N_{lqc,k} } \sum_{i: X_i = k} \ind{ R_i = lqc } \left( Y_i(1) - Y_i(0) \right) \nonumber \\
&= \sum_{k=1}^K \frac{N_{lqc,k}}{N_{lqc}} ITT_{lqc}^k \nonumber \\
&= \sum_{k=1}^K \frac{N_k}{N} \frac{ \pi_{lqc}^{k}}{\pi_{lqc}} ITT_{lqc}^k \label{eq:slice-sum}
\end{align}
The overall average treatment impact is a weighted average of slice-level treatment impacts, with weights depending on the size of the slice as well as the relative proportion of low-quality compliers in the slice to the proportion of low-quality compliers overall.
We are weighting each slice's average effect by the number of low-quality compliers in the slice.

Using the main results from above, we first obtain bounds within each of the slices:
\[ \mu_{lqc}^{k,low}(1) - \mu_{lqc}^k(0) \leq ITT_{lqc}^k \leq \mu_{lqc}^{k,high}(1) - \mu_{lqc}^k(0) \]
for $k = 1, \ldots, K$.

The overall bounds then immediately follow by plugging in the slice-specific lower (upper) bounds terms for the $ITT_{lqc}^k$ into Equation~\ref{eq:slice-sum} to get the overall lower (upper) bound. For example
\begin{equation}
 ITT_{lqc} \leq \sum_{k=1}^K \frac{N_k}{N} \frac{\pi_{lqc}^k}{\pi_{lqc}} ITT_{lqc}^{k,high} . \label{eq:ITT_high_sliced}
\end{equation}

The overall bounds from recombining the slices simply restricts the possible trade-offs of $ITT_{lqc}$ and $ITT_{hqc}$ as compared to the original full-sample bounds.
In particular, the overall trade-off of $ITT_{lqc}$ and $ITT_{hqc}$ is still linear---the overall weighted average of the treatment means still needs to average to $M_1$---so if we graph the potential set of treatment effect pairs, it will still be a line segment, and we can graph this set of possible pairs of $(ITT_{lqc},ITT_{hqc})$ by drawing a line connecting the two pairs of $(ITT_{lqc}^{low},ITT_{hqc}^{high})$ and $(ITT_{lqc}^{high},ITT_{hqc}^{low})$. This bound will be a subset of the original bound (up to estimation uncertainty). See the Supplementary Material for a simple argument for these observations.

\subsection{When stratification helps}

In examining the components of the overall bounds given on Equation~\ref{eq:bounds} and its high quality equivalent, we see that the width of the bound is primarily governed by the relative proportion of the two types of compliers, $\pi_{hqc}/\pi_{lqc}$ for low quality compliers and $\pi_{lqc}/\pi_{hqc}$ for high quality compliers.
There is a substantial trade-off here: more of one complier type means shorter bounds for that type and longer bounds for the other.
We can also achieve a narrower width if $M_1$ has an extreme value. For example, if $M_1$ is near 1, meaning the average outcome of all ECHS compliers is close to 1, this causes the bound to be truncated from above.  Similarly, if $M_1$ is close to 0, meaning the average outcome for all ECHS compliers is close to 0, this causes the bound to be truncated from below.

These observations suggest two strategies for obtaining narrow bounds within a slice.
The first, using the LQCs as an example, is to make slices with relatively more LQCs than HQCs to make our ratio for the LQC group small and the bound tight.
Second, if most of the students in a slice have the same outcome, then truncation could give a narrower bound.

The first strategy suggests we slice based on variables predictive of being a LQC versus a HQC.
One such variable would be the principal score, i.e., a prediction of what \emph{type} of complier a student would be, if they were a complier and given their full baseline covariate vectors.
We call related variables \emph{principal variables}.
The second suggests we slice based on variables predictive of the outcome.
One such variable is the \emph{prognostic score}, i.e., the predicted outcome of a student given his or her baseline covariates (and excluding treatment assignment).
We call related variables \emph{prognostic variables}.
One type of variable \emph{not} suggested by the above would be ones that predict overall ECHS compliance, i.e., those that predict being a complier (of any type) vs. an always-taker (which we call \emph{compliance variables}). 

One might worry that slicing to get tight bounds for one group would give poor bounds for the other, given the first strategy.
It turns out that this is not an issue: the first strategy essentially segregates the two types of units, and as the ratio for a group will tend to be large precisely for those slices that contain few students of that group, when we average the bounds across the slices, these wider bounds get low weight, producing tighter overall bounds.
In fact, even if in the overall sample one type dominates, as long as we can slice to have the less represented group mostly isolated in a few slices we can obtain tighter bounds for both.

We further explore these themes, as well as the question of whether to focus on principal or prognostic variables for slicing a sample, in a simulation study that we discuss in Section~\ref{sec:simulation}.



\section{Estimation}
\label{sec:estimation}
In the discussion above, the $\rho_{zs}$ and $\overline{Y}_{zs}$, are fixed, pre-randomization quantities (i.e., finite population parameters) and are not directly observable given the data.
We now turn to how to estimate these quantities with data and known random assignment mechanisms such as one has with a randomized experiment or school lottery.

Let estimates of the $\overline{Y}_{zs}$ be $\widehat{Y}_{zs}$ and $\rho_{zs}$ be $\hat{p}_{zs}$.
Then, as long as these individual estimates are asymptotically consistent, i.e., as long as $\hat{p}_{ze} \rightarrow \rho_{zs}$ and $\widehat{Y}_{zs} \rightarrow \overline{Y}_{zs}$ as $n$ increases, if we plug them in to our bounds formula the bounds estimates too will be asymptotically consistent.
Obtaining asymptotically consistent estimators is typically quite straightforward; in particular, weighted mean estimators tend to be consistent.

The simplest case is complete randomization.
Here each treatment arm is a random sample of the finite population, and the mean of any quantity of these samples will be an unbiased estimator of the mean for the whole finite population.
We would then take all our observed sample means and plug them into the bounds formulae.
Slightly more complex is when the probability of treatment varies for each student. Here we have to take these probabilities into account.
We discuss this below.

Generally, when using a plug-in approach, our direct estimates will be well behaved in that they will lie in the $[0,1]$ interval.
The bound estimates and estimated means can, however, lie outside this interval.
For example, the estimated proportion of LQCs or HQCs, because they are estimated as differences in observed proportions, could be negative: consider the case of randomly getting a higher estimated proportion of students in LQ schools under treatment than under control, which could easily occur if the proportion of LQCs was small.
We advocate simply truncating the proportions to the $[0,1]$ interval.
Overall, as long as our estimates are consistent, our bounds will be.

\subsection{Estimation with covariate stratification}

We illustrate estimation for a post-stratified bound estimate by walking through calculating the upper bound on the LQC group.
First, for each slice $k$, calculate $\hat{\pi}^k_{lqc}$ and $\widehat{ITT}_k^{lqc,high}$ using the weighted plug-in methods discussed below.
Next, and this is the subtle point, we calculate the estimated proportion of LQC in the entire sample as
\[ \hat{\pi}_{lqc} = \sum_k \frac{N_k}{N} \hat{\pi}^k_{lqc} .\]
This estimate comes from the relationship of
\[ \pi_{lqc} = \frac{1}{N} \sum_{i=1}^N  \ind{ R_i = lqc } = \frac{1}{N } \sum_{k=1}^K N_k  \frac{1}{N_k } \sum_{i: X_i = k} \ind{ R_i = lqc } = \sum_{k=1}^K \frac{N_k}{N} \pi_{lqc}^k . \]

We calculate the overall proportion this way (as opposed to taking the estimate from the raw, unsliced sample) so the slice weights sum to 1.\footnote{If we estimate our $\pi_{r}$ with our entire sample we will get different values.  The slice-averaging method is a form of post-stratification, so we are getting adjusted estimates that align with our overall averaging process.}
To obtain the overall bound, we would then plug these estimated values into Equation~\ref{eq:ITT_high_sliced}.
The other bounds are all analogous.

\subsection{Assessing uncertainty}
Once we have obtained bounds, we might wish to put uncertainty intervals on these bounds. 
Unfortunately, quantifying uncertainty for set identified parameters is technically challenging. 
See~\citet{Canay:2016wa} for a recent review.
We adopt a straightforward approach via the case-resampling bootstrap.
For the case-resampling bootstrap we (1) repeatedly resample the data and recalculate the bounds on the resampled data; and then (2) take the $5^{th}$ percentile of the lower bound and the $95^{th}$ percentile of the upper bound to get an overall adjusted 95\% confidence bound.\footnote{It may seem odd that we can merge two $1-\alpha$ level confidence intervals (rather than $1-\alpha/2$-level) to get an overall $1-\alpha$ coverage.
We can do this because of the bound itself.
Due to the bound, as long as the bound is suitably wide, we get proper coverage because if the parameter is close to a given extreme of the bound, the only possible error in coverage is one-sided.
In particular, if the parameter is at the low end, we only care about coverage of our lower interval.
If at the upper, only the upper interval.
This mutual exclusion means we do not need to ``split'' our $\alpha$ across both intervals.}

This adjusted bound is not a 95\% confidence interval for the full true bound, but it does have 95\% confidence for containing the true \emph{parameter}, as long as the estimation uncertainty is on the same order as or smaller than the true bound width. 
While current mathematical theory suggests that this bootstrap could break down in certain extreme cases, such as when the true parameters are at the limits of their bounds, we have found in further simulation work (not shown) that the bootstrap tends to work quite well in practice.
For additional discussion see \cite{yang2016using} or \cite{Bugni:2010fy}.
Furthermore, the bootstrap even performs admirably when we slice our data into a relatively large numbers of slices.
In particular, even though the estimated bounds within any given slice can be highly unstable, averaging across the slices produces shorter bounds that are as stable as the unadjusted bounds.

\subsection{Incorporating Weights}

In the ECHS study, a student's probability of treatment assignment varied by demographics and other factors.
In particular, some lotteries were more selective, and within some lotteries some students were given higher chances of a slot for equity reasons.
In such cases each treatment arm is a probability sample of the finite population, and we need to take the varying student-level probabilities of selection into account. 
Fortunately this is straightforward due to the known assignment mechanism.

Let $s_i = \pr{ Z_i = 1 }$ be the probability of student $i$ being selected into the treatment arm; then $1 - s_i$ is the probability for being in the control arm.
We then weight student $i$ by $w_i \equiv 1 / s_i$ if it is in the treatment arm and $w_i \equiv 1/(1-s_i)$ if control:
\[ w_i = Z_i \frac{1}{s_i} + (1-Z_i)\frac{1}{1-s_i} . \]

To illustrate estimation, consider estimating $\rho_{0e}$, the proportion of those in the control arm who enrolled in an ECHS school.
One simple estimator for $\rho_{0e}$ is
\[ \hat{p}_{0e} = \frac{1}{N} \sum_{i} \frac{1}{1 - s_i} (1-Z_i) \ind{ S_i = e } = \sum_{i : Z_i = 0} w_i  \ind{ S_i = e }, \]
as $w_i = 1/(1-s_i)$ for control students.
Because $\EE{ 1 - Z_i } = 1 - s_i$, the overall estimator is unbiased: $\EE{ \hat{p}_{0e} } = \rho_{0e}$.

The above, generally called a Horvitz-Thompson estimator \citep{Horvitz:1952tp} or the inverse probability of treatment weighted (IPTW) estimator, is the simplest.
It is unbiased, but is often suboptimal; Horvitz-Thompson estimators can be highly unstable.
In particular, it does not adjust for unlucky draws with imbalance in the number of  students with large weights.

The usual repair is to divide the sum not by the expected total weight $N$ but the total weight we actually sampled.
This is in fact the more natural weighted average of our sampled students.
It is slightly biased, but this bias is usually negligible.
Our final estimator, called a H\`{a}jek estimator, is then
\[ \hat{p}_{0e} = \frac{1}{Z} \sum_{i : Z_i = 0} w_i \ind{ S_i = e } \mbox{ with } Z = \sum_{i : Z_i = 0} w_i. \]
Similarly, our estimate of $\overline{Y}_{0e}$ would be
\[ \widehat{Y}_{0e} = \frac{1}{Z} \sum_{i : Z_i = 0} w_i Y_i^{obs}, \]
using the same normalizing constant $Z$.
See \cite{sarndal2003model} for a canonical overview of survey sampling methods.

All of these estimators are consistent: as sample size grows, they will converge on the true finite population values.
This means the associated bounds are consistent as well.
(We do not here discuss the finite sample asymptotics behind a formal argument; it is relatively straightforward.)
Importantly, the dependence of assignment patterns across students (e.g., from a block-randomized experiment) does not matter for the consistency.
This is important because it means when we slice our sample, we can still use our individual weights without modification, even if we slice across randomization blocks.
Finally, the weights used can be any weights to generalize from a sample to a population.
In particular, non-response or survey weights could be incorporated as above.


\section{Simulation Study}
\label{sec:simulation}

We simulate a series of fake data sets where students have three latent traits that govern their compliance behaviors and outcomes.
We use this complex model so we can generate data with different predictiveness in our covariates while keeping the same distribution of effects and outcomes in order to examine how different kinds of covariates improve bounds for the same data.

The three latent traits, all standard normal, are predictive of three different things:
\begin{enumerate}
\item $U_1$, a latent principal variable, determines whether the student is in a high or low quality school context.
Even if a student is an ECHS always-taker, he or she could still be in such a context.
\item $U_2$, the compliance variable, determines proclivity to go to an ECHS school. Those students with very high $U_2$ are ECHS always-takers, and those with low $U_2$ are always-takers of their school context from above.
\item $U_3$, the prognostic variable, is students' baseline ability, and predicts outcome (which is generated via a logistic model).
\end{enumerate}
  
So, for example,  a student with low $U_1, U_2, U_3$ would be a Low-Quality Always Taker, and would be more at risk of a zero for an outcome, relative to his or her peers.
A midrange value of $U_2$ would be a complier, and a high value of $U_2$ would be an ECHS always-taker.
This data generating process is parameterized by $\beta$, the predictive strength of $U_3$ on the outcome, $\omega_{r,z}$, the mean outcome for students in strata $r$ under treatment $z$, and the five strata proportions $\pi_r$.
Our observed covariates, $X_{ti}, t=1, 2, 3$, are noisy versions of the $U_{t}$:
\[ X_{ti} = U_{ti} + \epsilon_{ti} \mbox{ with } \epsilon_{ti} \sim N( 0, \sigma^2_{t} ) . \]
The greater the noise (determined by the $\sigma^2_t$), the weaker the connection between the covariates and the true structure of the data generating process.
Details of the data generation process, including how we connect our latent traits to observed school attendance behavior and outcomes, are in the Supplementary Material.


For our presented study, we calibrated our simulation by selected values for our baseline prognostic predictiveness $\beta$, our school effects $\omega_{r,z}$, and our proportions $\pi_{r}$ so the resulting data would closely match characteristics of the actual ECHS data (see \cite{kern2016assessing} for further discussion of calibration).
We made the $\overline{Y}_{zs}$, the marginal means of the six directly observable groups defined by treatment assignment and observed school type attended, as well as $M_0$ and $M_1$, the overall complier mean outcomes under control and treatment, match the ECHS estimates.
These choices ensured that our overall complier average causal effect was approximately 0.07, again matching the ECHS data.
Matching these moments still gives leeway, however, for selecting some of the $\omega$ values.
We tuned the final $\omega$ to give a slightly negative treatment impact to the HQC group and a large positive impact to the LQC group to examine cases where the true values were near the limits of the bounds.
We also matched our sample size of $N=3820$.
We also provide code in case readers wish to try different parameter combinations.

As a baseline we first examine how the unadjusted bounds performed by averaging the unadjusted  bound width across 30 no-noise datasets.
With no adjustment we found very short bounds (average width of 0.06) for the LQCs and much wider bounds (average of 0.39) for the HQC.
This comes from the LQC being such a large share (about 87\%) of the compliers.
The HQC bounds are still slightly informative, likely due to the generally high average outcomes overall, causing the bounds to be truncated from above at 1.

We can then imagine stratifying on any combination of our covariates $X_1$ (a principal-score style variable), $X_2$ (a variable predicting compliance), and $X_3$ (a variable predicting the outcome).
We can also vary the predictive power of our $X_{t}$ by adjusting $\sigma^2_t$ to see how strong our predictors need to be to see gains in our bounds.

\begin{figure}[hbt]
\center
\includegraphics[width=\textwidth]{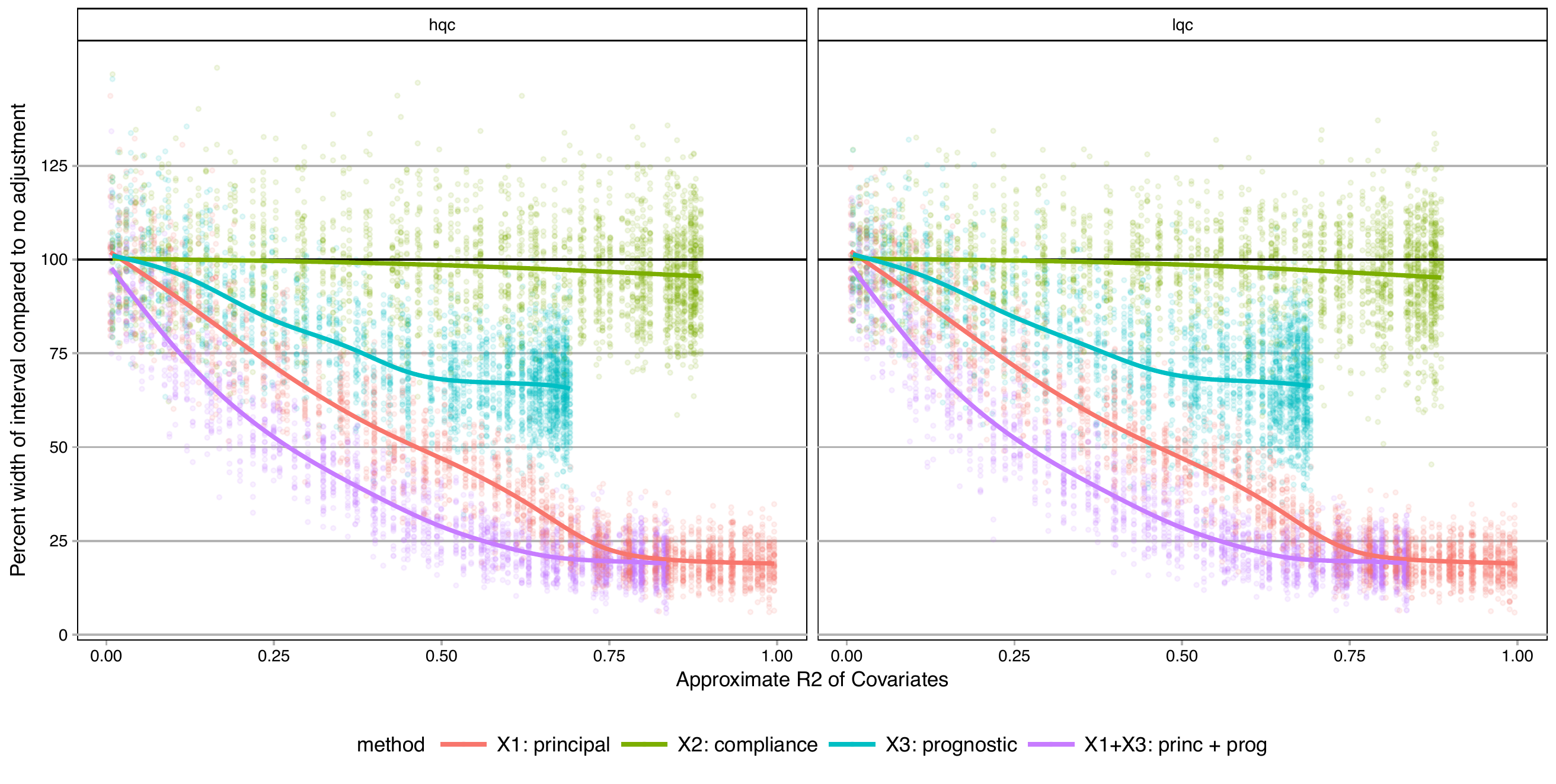}
  \caption[Bound width as a function of covariate strength]{Percent width of bounds (relative to no adjustment) as a function of the predictive power of covariates for simulation mimicking structure of ECHS data. Note that $R^2$ measures are only approximately comparable across covariates as the covariates are predicting different things.  Far right of lines correspond to no noise added to the original $U_{ti}$. }
  \label{fig:bound-performance}
\end{figure}

For each $\sigma^2$, we generated a single dataset with randomly assigned treatment and corresponding observed outcomes.
Then, for each covariate, we sliced our data into $K=4$ equal slices and estimated the overall bounds based on our synthetic data.
We also calculated an approximate $R^2$ measure of how predictive the used covariate was for its intended purpose to help comparison across variable types and help connect to applied practice.
(See Supplementary Material for details on these $R^2$ measures.)
Figure~\ref{fig:bound-performance} shows the resulting widths of the intervals, as a percent of the corresponding unadjusted width, for the ITTs for both the low and high quality compliers for different levels of noise.
There is variation in the individual points due to the natural uncertainty from the data generation process.
We smoothed the bound widths from the individual datasets and simulation trials with loess lines.
Several trends, discussed next, are evident.


\paragraph{Stratifying on $X_1$ (a principal variable).}
$X_1$ predicts the environment of the student, i.e., whether they would go to a high or low quality school if they did not go to an ECHS school.
This means High-Quality and Low-Quality compliers will tend to be in different strata.

We see that stratifying on $X_1$ can be highly useful: if we can form strata that are predominantly high- or low-quality compliers only, then we have good precision for those strata. When we average these strata-level bounds, we get good overall bounds as well. For example, in a strata dominated by HQC, we have tight bounds for HQC and loose ones for LQC.
For the strata dominated by LQC it is the reverse.
When we average, the loose bounds get low weight since they do not represent a large portion of the overall population.
See the Supplementary Material for further illustration of this.

\paragraph{Stratifying on $X_2$ (a compliance variable).}
Being able to predict whether a student is an always-taker or complier is not very useful for tightening bounds.
We see in Figure~\ref{fig:bound-performance} that the bounds tend to be about as wide as using no stratification, regardless of the predictive power of this covariate.
The reason stratifying on compliance is not useful is that, at least in this particular context, we can point identify the means of the always takers and subtract them out.
From an identification point of view, they therefore impact little.
This suggests that, more broadly, practitioners should focus on finding covariates that distinguish among those strata where quantities of interest are not point identified.


\paragraph{Stratifying on $X_3$ (a prognostic variable).}
In our considered scenario $U_3$ is fairly predictive of the outcome.
When slicing our sample by $X_3$ in the no-noise case, the average outcome for the control group is about 67\% in the first slice and 100\% for the last.
Slicing the sample on such a prognostic $X_3$ does tighten bounds for the HQCs, as is seen in Figure~\ref{fig:bound-performance} on the left hand side.
The gains primarily come from the extreme strata: when most students have 0s for their outcome or most have 1s, the greater homogeneity allows for tighter bounds.

\paragraph{Discussion.}
Overall, forming slices that are mostly one type of complier or the other is the key for getting tighter bounds under this framework.
Slicing based on a variable predictive of the outcome also provides gains if one can form groups that are close to the maximum or minimum outcome value possible.
One might naturally ask whether forming slices based on both principal and prognostic variables could be useful.
It can.
Because information of both of these scores is constraining the bounds by different methods, a combination does provide greater tightness (see the lowest line on Figure~\ref{fig:bound-performance}).
Unfortunately, slicing on both variables does create a large number of strata and estimation within those strata can be quite unstable.
Then again, averaging across these strata, as they are independent, could give the stability back.
Generally we found for our sample size of a few thousand that 12 slices, or even more, worked effectively; see supplement.
One might imagine further stabilizing this estimation with some sort of smoothing; we leave this as an important area of future investigation.

As a further exploration, we examined other parameter values.
In particular we examined cases where the overall mean outcome was not so close to 1 for so many of the strata.
The general trends found above replicated; see details and further commentary in the Supplementary Material.
Overall, slicing on both a principle-style and a prognostic-style covariate gives the best bounds.
The former is important because within a stratum the bound width, especially when the mean outcome is near 0.5, is governed by the ratio of complier proportions.
The latter is important for taking advantage of the truncation.

\section{Analysis of the ECHS dataset} 
\label{sec:application}

We return to the ECHS evaluation and examine how using different covariates reduced the width of our bounds.
These results were obtained by estimating the component proportions and means and plugging them in to the bounds formulae.
All calculations using the ECHS data make use of the ECHS sample weights that capture the students' different, known, probabilities of winning the lotteries.
These probabilities differed because some Early College High Schools gave priority to groups underrepresented in higher education, and because different sites had different levels of over-subscription.

\begin{table}[hbt]
\center
\small
\begin{tabular}{l|ccccc}
 &\multicolumn{1}{c}{ECHS }&\multicolumn{1}{c}{ Low Quality }&\multicolumn{1}{c}{High Quality }&\multicolumn{1}{c}{Low Quality }&\multicolumn{1}{c}{High Quality } \\
 &\multicolumn{1}{c}{ Always Takers}&\multicolumn{1}{c}{ Always Takers}&\multicolumn{1}{c}{ Always Takers}&\multicolumn{1}{c}{  Compliers}&\multicolumn{1}{c}{  Compliers} \\
&\multicolumn{1}{c}{(N=43)} & \multicolumn{1}{c}{(N=259)} & \multicolumn{1}{c}{(N=53)} & \multicolumn{1}{c}{(N=1328)} & \multicolumn{1}{c}{(N=220)}\\
 \hline
\multicolumn{5}{l}{Demographics} \\
 \hline
   American Indian&0.0\%&1.1\%&4.0\%&1.0\%&0.9\%\\
   Asian&0.0\%&1.0\%&0.0\%&1.0\%&3.3\% \\
   Black&23.5\%&34.0\%&23.5\%&29.0\%&15.0\% \\
   Hispanic&10.7\%&8.2\%&0.0\%&7.1\%&5.5\% \\
   Multi-racial&3.9\%&3.4\%&11.3\%&3.6\%&2.7\% \\
   White&61.9\%&52.3\%&61.2\%&58.4\%&72.7\% \\
   Male&45.1\%&38.7\%&30.1\%&40.1\%&39.2\% \\
 \hline
\multicolumn{5}{l}{Socioeconomic Background} \\
 \hline
   First Generation College  &27.6\%&46.2\%&36.1\%&41.5\%&41.4\% \\
   Free/Reduced Price Lunch &48.3\%&50.0\%&44.0\%&50.6\%&35.3\% \\
 \hline
\multicolumn{5}{l}{Exceptionality} \\
 \hline
Learning   Disabled &2.1\%&1.2\%&3.8\%&2.6\%&0.5\% \\
   Gifted&10.4\%&9.1\%&0.0\%&9.4\%&4.8\% \\
Ever Retained in grade&7.5\%&2.9\%&1.8\%&3.1\%&1.7\% \\
 \hline
\multicolumn{5}{l}{8th Grade Achievement} \\
 \hline
   Math - pass&74.3\%&80.6\%&87.2\%&80.8\%&97.0\% \\
   Reading - pass&77.0\%&78.2\%&85.9\%&79.5\%&92.5\%\\
   Algebra - take up&39.8\%&24.5\%&26.3\%&23.2\%&25.2\% \\
 \end{tabular}
  \caption{Summary statistics of the five principal strata}
  \label{tab:strat_description}
\end{table}

\begin{figure}[hbt]
\center
\includegraphics[width=0.9\textwidth]{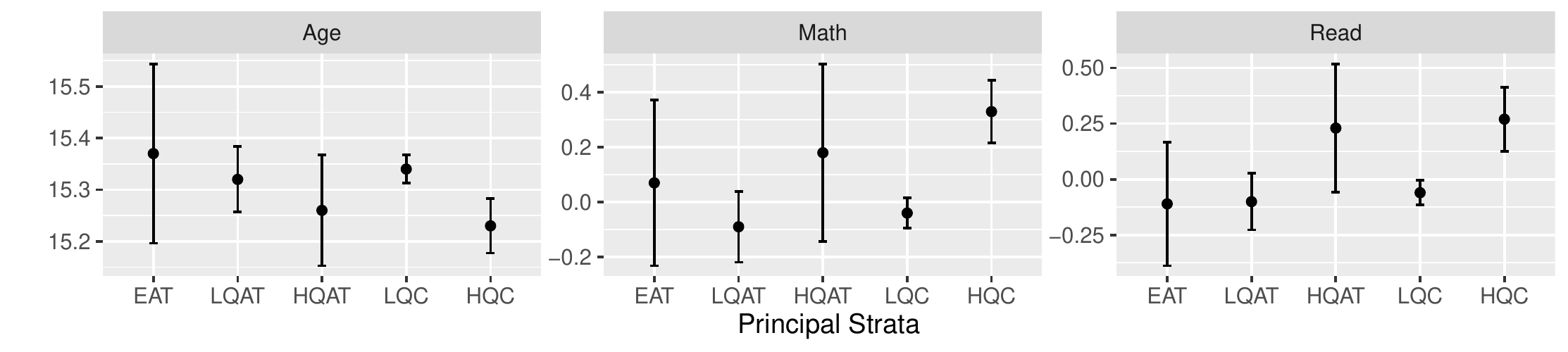}
  \caption[Continuous covariate means by strata]{Means of the continuous pre-treatment covariates by strata for the ECHS data with approximate 95\% confidence intervals.}
  \label{fig:covariate-plot}
\end{figure}

First, using a strategy analogous to calculating the relative sizes of the strata, we estimated the means of different baseline demographic and achievement covariates of these strata to further characterize these different populations.
Results are in Table~\ref{tab:strat_description} and Figure~\ref{fig:covariate-plot}.
The 8th grade scores are standardized by year to account for different scales in years students were tested.
The baseline characteristics of students across our five strata vary substantially; thus differential impacts across these groups may be due to differences in characteristics rather than differences in the type of high school they attended.
For example, among compliers who attended high quality schools, 35\% were eligible for free lunch, substantially less than the other groups, which implies that the high quality schools available to students participating in lotteries may have been located in higher income areas.
These students also outperformed students in other strata in 8th grade: 97\% passed the Math End-of-Grade exam in 8th grade and 93\% passed the English exam.
Passing rates for students in other strata range from 74\% to 87\%.
However, despite these differences, a similar proportion of high quality compliers and low quality compliers would be first-generation college students if they attended college.

Those who would always attend their non-ECHS school regardless of ECHS offer (the always-takers) appear similar to their complier counterparts.
The low quality always-takers, for example, have similarly low pass rates to the low quality compliers.
The high quality always-takers also appear similar to the high quality compliers, although their small numbers makes inference difficult.



The first and fifth rows of Table~\ref{tab:ECHS_bounds_nocovs_covs}, summarizing Table~\ref{tab:identified_estimates}, present the results of bounding without using baseline covariates.
Under assignment to control, 86\% of low-quality compliers are on-track at the end of 9th grade. Under assignment to treatment, the bounds indicate that between 94\% and 100\% of the low-quality-compliers are on-track. This translates to an ITT effect for low-quality compliers of between 8.1 and 13.7 percentage points.
Even at the lower bound, this effect is larger than the overall ITT estimate.
For the high-quality compliers, we estimate that nearly all (98.6\%) are on track at the end of 9th grade under assignment to control, and our bounds indicate that between 65\% and 100\% of high quality compliers are on-track under assignment to treatment.
Thus, we bound the treatment effect for high-quality compliers to be between -34\% and 1\%.
The impact associated with attending an ECHS appears to be much less pronounced and possibly negative for students who otherwise would attend a high quality high school, most likely owing to the strong outcomes these students experience when they are not offered the ECHS opportunity.
That is, because nearly all of these students would be on-track at the end of ninth grade without the ECHS opportunity, there is little room for improvement in their on-track status.
The bounds on the treatment effect that we estimate for high-quality compliers is quite wide mainly because this group makes up a small share of the overall sample.

We then employed several strategies to tighten the bounds on the treatment impacts for the two complier groups.
These results are also summarized in Table~\ref{tab:ECHS_bounds_nocovs_covs}.
We first utilized a baseline measure of student achievement.
Specifically, we subdivided the sample into quartiles based on performance on the North Carolina 8th grade End-of-Grade math assessment.\footnote{8th grade test scores are not available for all students. 11\% of students in our sample are missing 8th grade test scores. Students who applied to an ECHS lottery, but did not attend a North Carolina public school will not have 8th grade test score data. These students make up 20\% of the missing observations. Other missingness can be attributed to lapses in administrative data. We used regression imputed scores for students who were missing test scores, using pre-treatment covariates only, and not using treatment assignment.  Importantly, the independence of the treatment assignment to the imputation step means imputation under misspecification would not impact the validity of the bounds.}
Middle school math scores are prognostic, as they are predictive of whether a student will be on-track in 9th grade.
Our $R^2$ measure for this variable was 0.11.
We calculated the bounds for each quartile of students and averaged the results together, weighting by the number of students in each principal stratum in each quartile.
The bounds on both the high-quality and low-quality compliers are tightened by 32\% through this stratification method, much more than we expected given our simulation study.

We then grouped students based on the school they attended in 8th grade (the year in which they applied to the ECHS).
We chose 8th grade school based on the notion that students' 8th grade schools are highly predictive of the high school they would attend, given predominant school feeder patterns.
Sorting students on a characteristic that predicts principal stratum membership will create more homogenous strata and produce tighter bounds.
We gave each 8th grade school a score for each year by calculating the percentage of students in the ECHS study from that school who subsequently attended a high quality high school.
High school quality is measured for the year the student was in 8th grade, so students do not contribute to the quality rating of their chosen 9th grade school.
The $R^2$ measure here was 0.6; this variable is highly predictive and so we would expect great gains in our bounds, given the simulation study.
Using this measure, we split the sample into quartiles.\footnote{We had one technicality here. Several 8th grade schools had tied scores of 0 for some of their years so we broke ties using the overall percentage of students across the years of the study who attended a high quality high school (the quality metrics of some high-schools changed across time) so we could have well-defined quartiles of students.}
Finally, we calculated the bounds for each of the quartiles and combined the results, weighting for the number of students in each stratum and quartile.
This process resulted in narrower bounds on treatment effects for both the high-quality compliers and low-quality compliers as compared to the prognostic math score approach.

Our third approach was to stratify students on both their 8th grade achievement and their 8th grade schools.
Within each quartile based on middle school, we sorted students into three equally size groups based on their standardized 8th grade math test score, calculated the strata-level bounds in the resulting 12 slices, and averaged appropriately.
This results in a narrower bound than when we just use one or the other of our covariates.
For the HQC group, we bound the treatment impact between -8 and 3 percentage points, which is about a 69\% smaller range than the results of the bounding exercise without stratification.
Conceptually, we have generated slices that separate, as much as possible, the LQC and HQC students.
Then, by further slicing to group students likely to be on track, we created relatively homogenous groups which gives us narrower bounds by clipping the upper bound by 1.
This provided both benefits, giving the tightest bounds.

\begin{figure}[hbt]
\center
\includegraphics[width=0.9\textwidth]{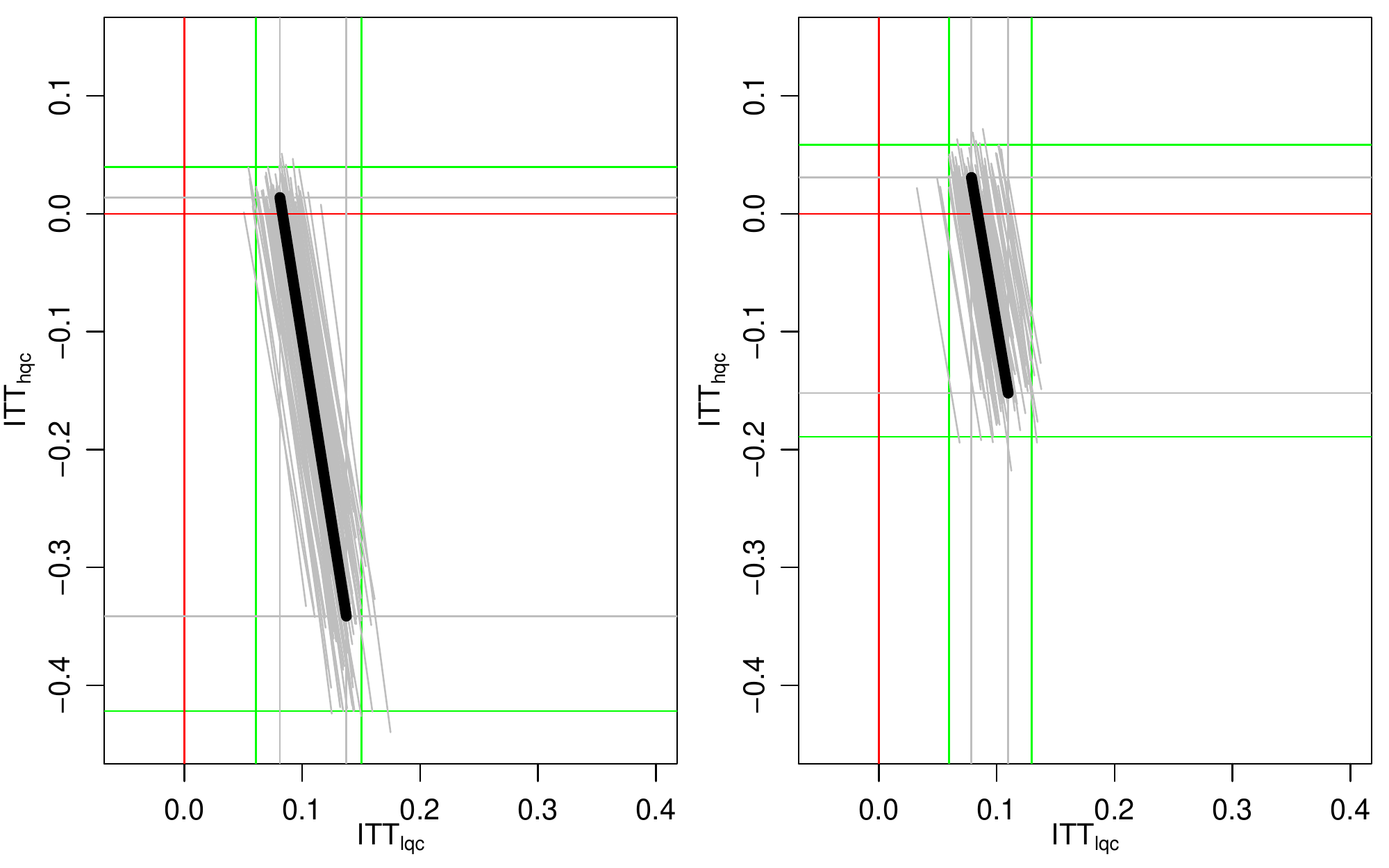}
  \caption[Bootstrap adjusted bounds for ECHS data with plausible set of treatment effects]{Bootstrap adjusted bounds for ECHS data with plausible sets of treatment effects.  Left hand side is raw bounds, right hand side is bounds from using both middle school and math scores.  Green lines indicate bounds incorporating estimation uncertainty.}
  \label{fig:boot-bounds}
\end{figure}

We finally incorporated estimation uncertainty with the bootstrap.
The right three columns of Table~\ref{tab:ECHS_bounds_nocovs_covs} show the final impact bounds adjusted for this uncertainty.
The estimation uncertainty expands the bounds rather substantially, but as the LQC and HQC intervals do not overlap, we have good evidence that the impact for the LQC group is higher than the HQC group.
We can see an illustration of the bootstrap on Figure~\ref{fig:boot-bounds}, where we plot 100 plausible treatment impact sets for 100 bootstrap draws.
The final bounds are formed by dropping the 5\% highest and lowest extremes from these sets. (See discussion of estimation for why 5\% and not 2.5\%.)

Both Figure~\ref{fig:trade-off-figure}(b) and the bootstrap adjusted version Figure~\ref{fig:boot-bounds} allow exploration of the possible trade-offs between impact on the LQC group and the HQC group.
For Figure~\ref{fig:boot-bounds}, any point in the central mass of grey lines is plausible.
We see that in order for the impact on LQC to be on the larger end of the scale, the impact on the HQC group would have to be substantially negative.
While this would be consistent with the data, one might argue from a substantive standpoint that the near zero HQC effects coupled with the lower end of the LQC effects (of about 8 percentage points) is most plausible.
We finally note that, as being at the absolute extreme ends of our bounds is substantively implausible here, we have good confidence in the bootstrap.

\begin{table}[ht]
\centering
\begin{tabular}{lrccrr|cr}
  \hline
 Stratification &	Co    &	Tx & ITT	&	Bounds      & \%  	   &  Adj.    &	Final \\
               &  Mean & Mean & Bounds & width & width & Bounds &  Width    \\
\hline
 \multicolumn{8}{l}{High Quality Compliers} \\
   \hline
None & 98.6 & 64.5 - 100.0 & -34.1 - 1.4 & 35.5 & 100 & -43.4 - 3.8 & 47.2  \\
  Math & 98.0 & 75.9 - 100.0 & -22.1 - 2.0 & 24.1 & 68 & -28.9 - 4.9 & 33.8  \\
  Middle School & 97.6 & 84.1 - 100.0 & -13.5 - 2.4 & 15.9 & 45 & -19.0 - 4.5 & 23.5  \\
  Both & 97.5 & 89.1 - 100.0 & -8.4 - 2.5 & 10.9 & 31 & -13.2 - 5.1 & 18.4  \\
   \hline
 \multicolumn{8}{l}{Low Quality Compliers} \\
   \hline
None & 86.3 & 94.4 - 100.0 & 8.1 - 13.7 & 5.6 & 100 & 5.7 - 15.3 & 9.5  \\
  Math & 86.4 & 94.3 - 98.1 & 7.9 - 11.7 & 3.8 & 68 & 6.1 - 13.8 & 7.7  \\
  Middle School & 86.5 & 94.0 - 97.4 & 7.5 - 10.8 & 3.4 & 59 & 5.3 - 12.8 & 7.6  \\
  Both & 86.6 & 94.0 - 96.3 & 7.4 - 9.7 & 2.3 & 40 & 5.4 - 12.1 & 6.7  \\
   \hline

   \end{tabular}

  \caption[ECHS treatment effect bounds with and without covariate stratification]{ECHS treatment effect bounds with and without covariate stratification.  All outcomes are in percents.  ``Percent width'' compares bound width to width when no covariates are used. Final columns display bootstrap adjusted bounds and widths that takes into account estimation uncertainty.}
  \label{tab:ECHS_bounds_nocovs_covs}
\end{table}

\section{Conclusions}

Estimating treatment effects for subgroups defined by post-treatment behaviors can be challenging, as we can only observe some of these groups as mixtures.
The researcher then typically has a choice: impose assumptions which allow for direct estimation (e.g., conditional independence assumptions of strata membership and outcome conditional on pre-treatment covariates), model the relationship of the covariates and outcomes in order to separate the mixtures, or calculate bounds to obtain a range of plausible estimates that are consistent with the data.

In this work, we explore this last choice, demonstrating how to calculate such bounds in the context of an experiment where we aim to separate the overall complier average causal effect according to the behaviors that compliers would exhibit when assigned to control.
We show that these bounds can be informative, producing a description of treatment impacts more nuanced than an overall average treatment effect.
We also demonstrate how such bounds can be tightened using covariates either predictive of outcome or stratum membership.
In this context, we find that covariates predicting stratum membership tends to lead to tighter bounds, but that one can tighten bounds even further by using both types of covariates, thus generating subgroups that are as homogenous as possible in terms of both stratum membership and the outcome.

In our evaluation of ECHS we used single variables that predicted outcome and compliance type.
As an alternative, one could regress the outcome (or school type attended) on the full range of covariates in order to make prognostic or principal scores, and stratify based on those scores, which could give additional gains.
This approach might be important when one has no single, strongly predictive covariate to use.
For one example of a similar approach see \cite{chan2017partially}, who uses bounds to generalize experiments.
Here, even if the model were misspecified, this process would give valid bounds as the scores would be pre-treatment quantities, provided the model fitting were not dependent on the treatment assignment. 
This independence can be achieved by sample-splitting (see, e.g., \cite{abadie2013endogenous}).
We leave this for future work.

When taken as a whole, we believe that, when designing studies, practitioners should spend effort on collecting covariates that not only strongly predict the outcome (which is typically done with, e.g., pre-tests) but also that predict the post-treatment behaviors of interest.
We have found in both our simulations and applied example that stratifying on both these types of variables gives the strongest gains.
Stratifying on principal variables predictive of behavior is critical; without them the bounds can be quite wide.
Despite the risk of wide, uninformative bounds, we nevertheless advocate for using them, as the additional assumptions typically necessary for point estimation are generally quite strong and often implausible.
See, for example, \cite{Feller:2017wj}.

For our particular ECHS study, we found the estimated treatment impact on students who would attend a low-quality high school is consistently larger than the impact on the whole sample, while the range of impacts on students who would otherwise attend a high quality school includes 0.
This result comes from the systematically high achievement of the students attending the high quality schools; because they are predominantly on track, there is little room for improvement.

Broadly, these findings point to the importance of attending to context when interpreting program impacts \citep{Lemons:2014kk}. Specific to the ECHS context, these findings suggest that early college high schools may be a particularly strong intervention for those communities in which students have little access to high quality high schools. Policy makers looking to support educational attainment for these students may look to the early college model as a potentially positive intervention.

\bibliographystyle{abbrvnat}
\bibliography{references}

\newpage
\setcounter{page}{1}
\begin{center}
\title{\huge Supplementary Material\\for\\``Bounding, an accessible method for estimating principal causal effects, examined and explained''}
\end{center}

    \setcounter{equation}{0}
        \setcounter{section}{0}

\section{Data generation process for the simulation study}

We parameterize our data generating process with the proportions of the five strata, the predictive power of $U_3$, and the impact of the three school types on student outcomes.
In formal terms, our model is
\begin{align*}
U_{1i}, U_{2i}, U_{3i} &\sim 	N( 0, 1 ) \mbox{ (all independent) } \\
H_i &= \begin{cases}
      hq & U_{1i} > \Phi^{-1}\left( \gamma_0 \right) \\
      lq & else
   \end{cases}\\
S_i(0) &= \begin{cases}
      e & U_{2i} > \Phi^{-1}({1-\pi_{eat}}) \\
      H_i & else
   \end{cases}\\
S_i(1) &= \begin{cases}
      e & U_{2i} > \Phi^{-1}\left( \gamma_1(H_i) \right) \\
      H_i & else
   \end{cases} \\
\pr{ Y_i(z) = 1 } &= logit^{-1}\left( \beta U_3 + \omega_{R_i,z} \right)
\end{align*}
with $\Phi(\cdot)$ being the normal CDF, the $Y_i$ being Bernoulli coin flips with the given probabilities, and with
\begin{align*}
 \gamma_0 &= 1 - \frac{\pi_{hqc}+\pi_{hqat}}{1-\pi_{eat}} \\
 \gamma_1(hq) &= \frac{\pi_{hqc}}{\pi_{hqc}+\pi_{hqat}} (1 - \pi_{eat} ) + \pi_{eat} \\
 \gamma_1(lq) &= \frac{\pi_{lqc}}{\pi_{lqc}+\pi_{lqat}} (1 - \pi_{eat} ) + \pi_{eat} \\
\end{align*}
for a given, pre-specified collection of strata proportions.
For interpretation, $\gamma_0$ is the relative proportion of students in a low-quality context, ignoring the ECHS always-takers.
The $\gamma_1$ are cut-offs for going to an ECHS school under treatment, and will generally be less than $\gamma_0$ if we have positive proportions of compliers.

The $\beta$ is the slope connecting $U_3$ to $Y$, with high values making $U_3$ more discriminatory.
The $\omega_{r,z}$ are the mean impacts for each strata under treatment and control (so $\omega_{eat,0}=\omega_{eat,1}$, etc., for the always-takers).
These quantities shift the probability of on-track by some set amount on the logistic scale.
The $\omega_{lqc,1} - \omega_{lqc,0}$ represents the impact of treatment on LQCs going from a LQ school to an ECHS school.
This model allows for heterogeneous effects for the different compliers.

For our ECHS simulation we set our $\beta$ to 5.65, our $\omega_{r,0}$ to $17, 8, 13, 6, 6.5$ and our $\omega_{lqc,1}$ to 9.6 and $\omega_{hqc,1}$ to 11.
The $\pi$s were set to $0.03, 0.03, 0.11, 0.11, 0.72$.
58\% of the units were randomized to treatment.
For the primary plots we systematically varied the $\sigma^2_1 = \sigma^2_2 = \sigma^2_3 = \sigma^2$ from to attempt to achieve  from 0 to 3.

\subsection{Calculating $R^2$}
For each $X_{t}$ variable we calculated an $R^2$ that could be estimated from the observed data.
We wanted to use $R^2$ measures that practitioners could use to assess their actually available covariates in practice.
For each measure we regressed an observed binary outcome or behavior onto the covariate using a logistic regression.
We then compared the fitted model to a null model $\mathcal{M}_0$ using McFadden's $R^2$:
\[ R^2 = 1 - \log L( \mathcal{M} ) / \log L( \mathcal{M}_0 ) , \]
where $L( \mathcal{M} )$ is the likelihood of model $\mathcal{M}$ given data.

The models were as follows:
\begin{itemize}
	\item For a principle-style variable $X_1$ we regressed attendance of a low-quality school onto $X_{1}$ on the subset of control-group students not observed to be in an ECHS school.
	\item For a compliance-style variable $X_2$ we regressed attendance at an ECHS school under the treatment condition onto $X_{2}$ for the treatment group. This will be reasonable provided the proportion of ECHS always-takers is not too high.
	\item For a prognostic-style variable $X_3$ we regressed outcome onto $X_3$ and school type attended for the control group. Here the null model is the outcome regressed on just school type attendance dummies.
\end{itemize}

These $R^2$ measures ideally indicate the the strength of the covariates relationship to the underlying structures of interest.
The simulations can then be used as a diagnostic by comparing actual estimated $R^2$ measures found in ones data to the simulation findings.
The ability to compare covariates is admittedly not perfect.
In particular, for the prognostic covariate, the $R^2$ measure is typically low even in the no-noise case, due to the mixture of different principal strata.

\section{Characteristics of the plausible set of effects under slicing}
We here argue that, when we stratify on a covariate and then combine to get overall bounds, we still get a linear constraint between the impact on the LQC and the impact on the HQC groups regardless of how we slice.
In particular, the set of possible pairs $ITT_{hqc}, ITT_{lqc}$ will lie on a line.
Furthermore, up to estimation error, this line will be the same line as the possible pairs obtained from calculating bounds on the full dataset without slicing.

To see this first note that the overall average control-side means must be the same as the non-sliced case as they are all point identified.
Similarly for the overall proportions of the five principal strata.
We then focus on the aggregate overall treatment means.
The weighted average of these means, weighted by the relative proportions $\pi_{hqc}$ and $\pi_{lqc}$ must equal $M_1$.
But these pairs of means are precisely those that lie on the line of possible mean pairs, which induces the line of possible treatment effect pairs.

This means that even after slicing, we can return the set of possible pairs of effects as all points on the line segment between the two extremes of $(ITT_{lqc}^{low},ITT_{hqc}^{high})$ and $(ITT_{lqc}^{high},ITT_{hqc}^{low})$.

These equalities are on the population parameters.
Because we are actually estimating these quantities, and because the slicing is actually a form of post-stratification, giving adjusted estimates of the overall means and proportions, the set of solutions from a sliced bound will not necessarily be on the exact same line as the overall, although they should be close.
They would align with overall estimated bounds calculated from the post-stratified overall mean outcomes and proportions, however.

\section{Illustration of different bound precisions across slices}

\begin{figure}[hbt]
\center
\includegraphics[width=0.8\textwidth]{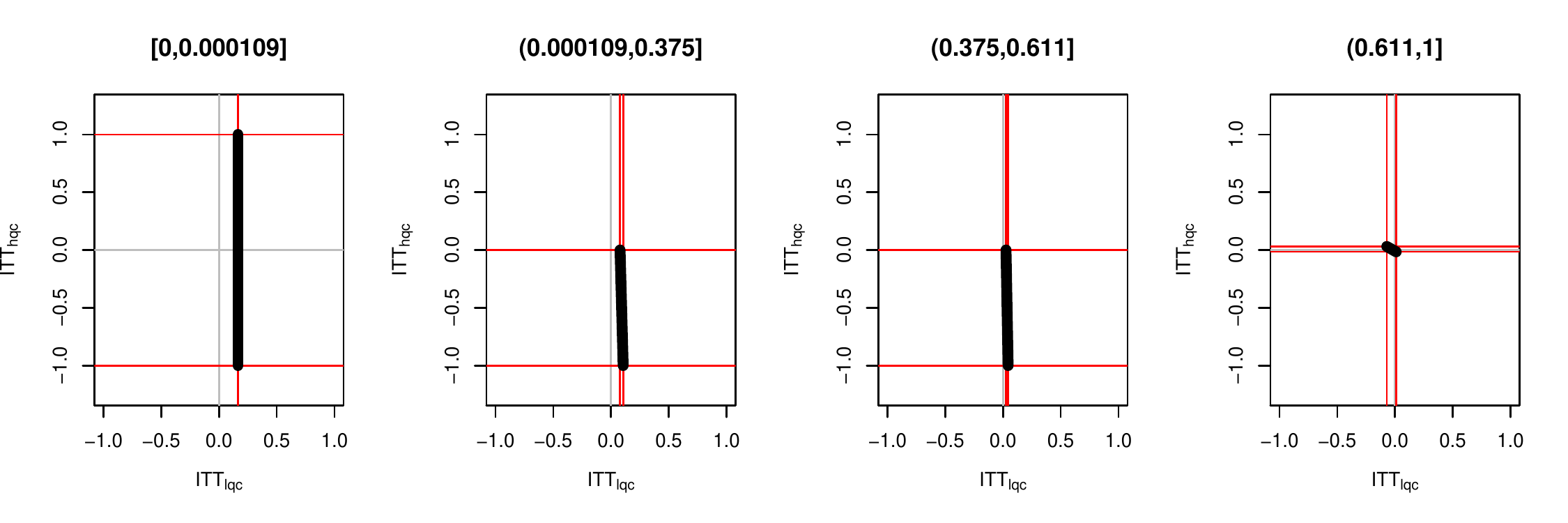}
  \caption[Variation of bounds across slices based on principal variable]{Variation of bounds across slices of ECHS data when slicing on the principal variable of percent attendance to a high-quality school.  Far left stratum has no students in the control going to a high-quality school, so bounds are $[-1,1]$ for the HQC. Middle strata are primarily LQC, so the HQC bounds are still uninformative. By contrast, the final, far right stratum has tight HQC bounds due to both being a high proportion of HQC and having high proportion of 1s for outcomes. The LQC bounds are here the widest, on the other hand.}
  \label{fig:strat-demo-plot}
\end{figure}

Figure~\ref{fig:strat-demo-plot} shows the individual strata-level bounds when stratifying on a variable predictive of strata type (percent high quality school attendance) for the ECHS data.
The first slice has only LQC and so we have a single point for our bounds for the LQCs and no contribution for the HQCs.
The second slice gives tight bounds for the LQC and loose ones for HQC, but this stratum is dominated by LQC, so it has high weight for the LQC bound when we average.
The loose HQC bound gets low weight and so the lack of precision does not matter much.
This is further illustrated on Table~\ref{tab:weight-chart}, which shows how the slicing makes slices with more $LQC$ or more $HQC$, which changes the bound width for each group.
In particular, see how as the weight of the LQC strata for each slice goes down the ratio governing the bound width for this group goes up.

\begin{table}[ht]
\centering
\begin{tabular}{lrrrrr|llr}
 	 B & eat & hqat & hqc & lqat & lqc & $\pi_{hqc}$ & $\pi_{lqc}$ & $\pi_{hqc}/\pi_{lqc}$ \\
  \hline
  [0,0.000109] & 0.28 & 0.00 & 0.00 & 0.26 & 0.32 & 0\% & 85\% & 0.00 \\
  (0.000109,0.375] & 0.46 & 0.08 & 0.03 & 0.35 & 0.26 & 2\% & 75\% & 0.03 \\
  (0.375,0.611] & 0.12 & 0.30 & 0.03 & 0.25 & 0.30 & 2\% & 83\% & 0.02 \\
  (0.611,1] & 0.14 & 0.62 & 0.94 & 0.14 & 0.11 & 55\% & 31\% & 1.76 \\
  TOTAL & 1.00 & 1.00 & 1.00 & 1.00 & 1.00 & 14\% & 69\% & 0.21 \\
   \hline
\end{tabular}
\caption[Relative proportions of strata within slices defined by quantiles of middle school score]{Relative proportions of strata within slices defined by quantiles of the principal variable of the middle school feeding patterns. Final three columns shows the proportion of HQC and LQC in the slice and the ratio of the LQC and HQC proportions, which governs bound width for the LQC group.}
\label{tab:weight-chart}
\end{table}

\begin{figure}[hbt]
\center
\includegraphics[width=0.8\textwidth]{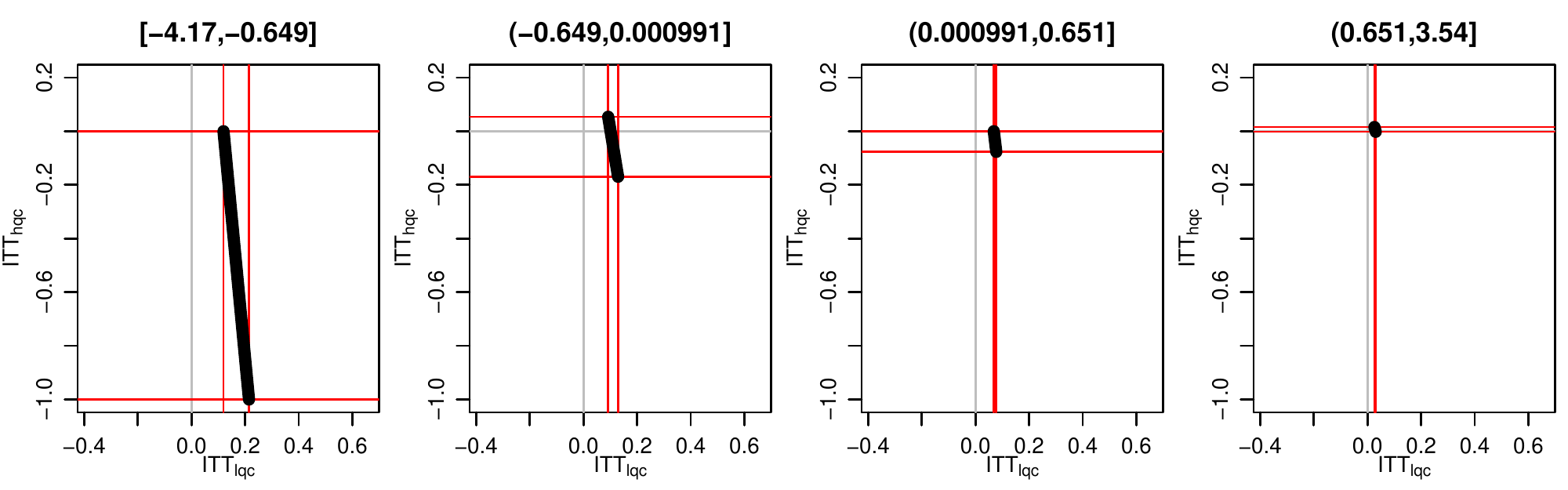}
  \caption[Variation of bounds across slices based on prognostic variable]{Variation of bounds across slices of ECHS data when slicing on the prognostic variable of 8th grade math.  Far left we see wide bounds for both LQC and HQC due to being far from the extreme of 1. Far right the bounds are tight as virtually all students have an outcome of 1, leaving little room for any impact.}
  \label{fig:strat-demo-plot-prognostic}
\end{figure}

For the prognostic-style variable with the ECHS study we used 8th grade math. When we stratify on a standardized version of this variable we see that for both the treatment and control groups the mean outcomes generally increase.
See Table~\ref{tab:echs_prognostic_summary}.
In particular, for the fourth stratum, virtually all of both treatment and control students are on-track.
This constraint gives the bounds as shown on Figure~\ref{fig:strat-demo-plot-prognostic}.
When we average the very wide bounds and the much narrower bounds across the strata, we obtained shorter overall bounds.

As a side-note, the ITT impacts vary by this covariate substantially.
In particular, we estimate a 9 percentage point gain for the weakest stratum, and only a 2 percentage point gain for the strongest.

\begin{table}[ht]
\centering
\begin{tabular}{lllll}
  B & \% (Co) & \% (Tx) & $\widehat{Y}_0$ & $\widehat{Y}_1$ \\
  \hline
   [-4.17,-0.649] & 26\% & 25\% & 73\% & 82\% \\
  (-0.649,0.000991] & 23\% & 27\% & 88\% & 96\% \\
   (0.000991,0.651] & 24\% & 24\% & 94\% & 98\% \\
   (0.651,3.54] & 26\% & 24\% & 97\% & 99\% \\
   \hline
\end{tabular}
\caption{Mean outcomes for ECHS data for slices defined by 8th grade math score for both treatment and control groups.  Middle columns denote proportion of sample in each slice for treatment and control sides.}
\label{tab:echs_prognostic_summary}
\end{table}

\section{Further simulations}

\begin{figure}[hbt]
\center
\includegraphics[width=0.8\textwidth]{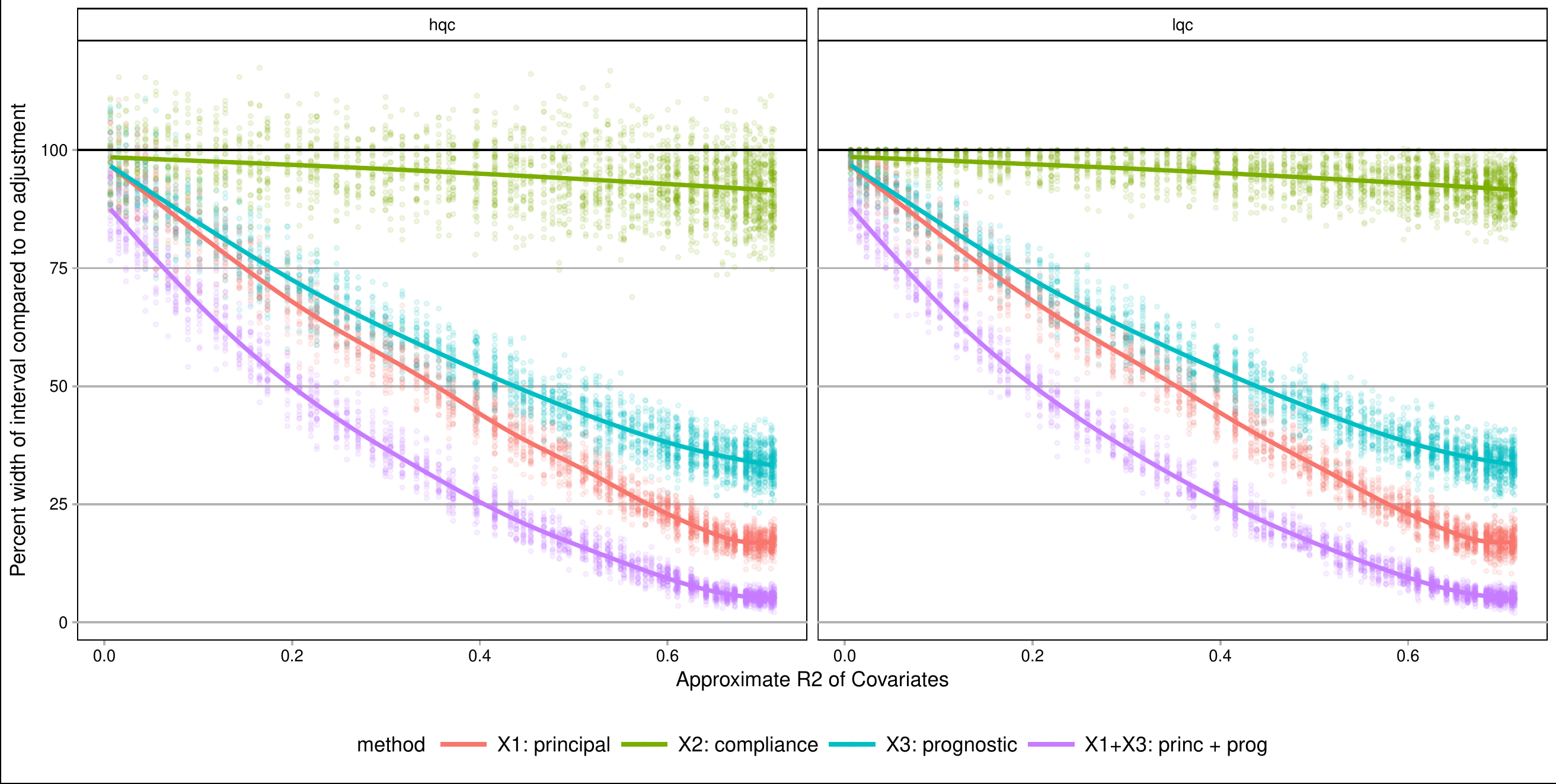}
  \caption[Extra simulation results]{Percent width of bounds (relative to no adjustment) as a function of the predictive power of covariates for alternative simulation scenario.}
  \label{fig:bound-performance-additional}
\end{figure}

We further examined bound performance by exploring other parameter values in our Data Generating Process.
Figure~\ref{fig:bound-performance-additional} shows a case where the two groups are more balanced and the mean outcomes are lower.
Parameters are set so the overall CACE is $0.12$, with a mean complier outcome under control of 0.47 and under treatment of 0.59.
We also sliced into 6 slices rather than 4.

Here, with $\beta = 4$, $X_3$ is highly predictive of the outcome in the no-noise setting.
For this setting, the average outcome for the control group in the first slice is about 1\%, and for the last it is 99\%. ($X_3$ has an $R^2$ of 0.58.)
Slicing the sample on such a prognostic $X_3$ substantially tightens bounds, as is seen in Figure~\ref{fig:bound-performance-additional}: the bound is about a third of the raw bound width for the $\sigma^2_3=0$ case (see the right endpoint of the trend line).
The gains primarily come from the extreme strata: when most students have 0s for their outcome or most have 1s, the greater homogeneity gives tighter bounds.
For further illustration see prior section investigating bounds within the ECHS slices.

Overall, in this scenario the benefits of stratification are even more striking than for the ECHS scenario, although bounds are also systematically much wider.
This simulation further shows that having outcomes very close to 1 (or 0) overall is not necessary for informative bounds, provided that one has reasonably predictive covariates.

\paragraph{Compliance Variables.}
Note that in the auxiliary simulation slicing on the compliance variable is beneficial when the variable is highly predictive (the low noise case).
This gain comes from a side effect: due to the imbalance in size of the LQC and HQC groups, of one of the slices having many HQCs and no LQCs.

\subsection{Examining sample size}
\begin{figure}[hbt]
\center
\includegraphics[width=0.8\textwidth]{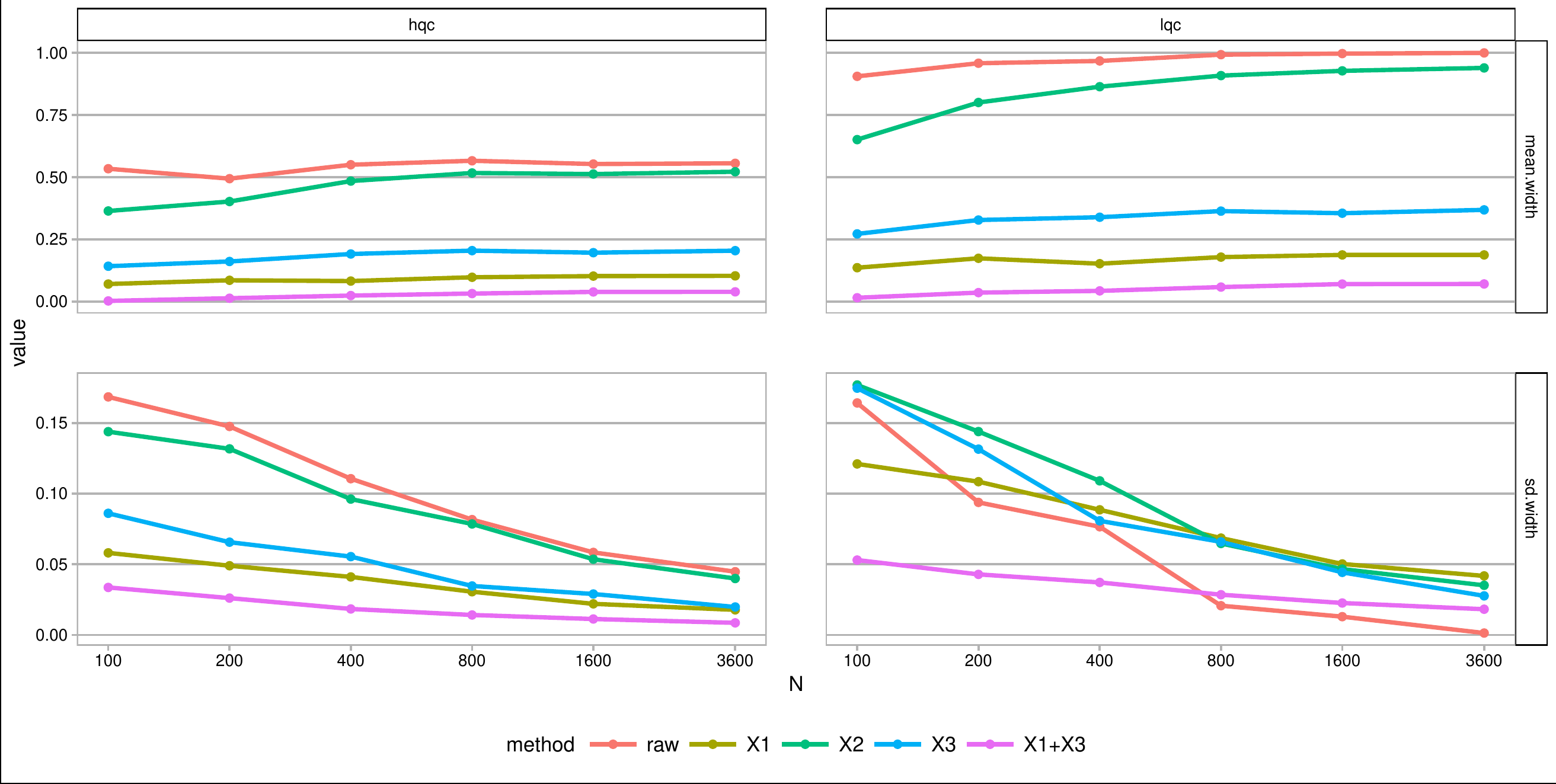}
  \caption[Extra simulation results on sample size]{Average width and variability of width of bounds as function of sample size.}
  \label{fig:sample-size-impact}
\end{figure}

To illustrate the impact of varying sample size, we re-ran the auxillary simulation with different sample sizes ranging from $N=100$ to $N=3600$.
Figure~\ref{fig:sample-size-impact} shows the results of 100 random trials at each sample size.
We see the average width of the bounds is stable, unrelated to sample size.
The variability of the bound width, however, goes down with sample size as expected.
In this scenario, for the methods with reasonably informative bounds, variability of sample size is on the same order as bound width for the larger sample sizes.
This suggests adjustment of bounds to account for uncertainty is generally a real concern.

\paragraph{Examining number of slices}

\begin{figure}[hbt]
\center
\includegraphics[width=0.8\textwidth]{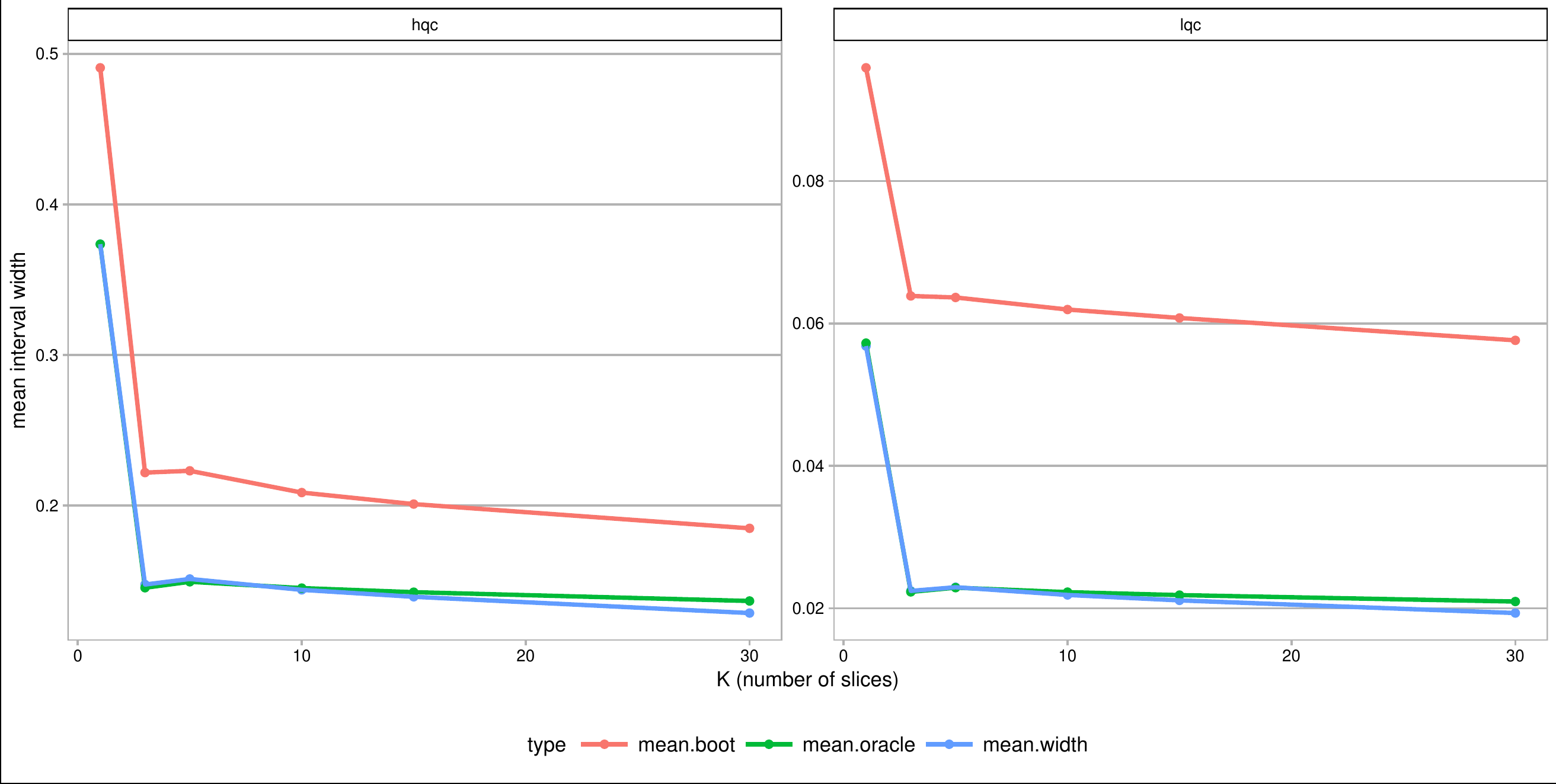}
  \caption[Extra simulation results on number of slices]{Average bound width, oracle bound width (where all identifable parameters are known), and bootstrap-adjusted bound width as a function of number of slices for propensity variable $X_1$ with $R^2 \approx 0.5$ for ECHS simulation.}
  \label{fig:num-slice-impact}
\end{figure}

We ran a final simulation to briefly investigate bootstrap coverage and the impact of the number of slices (we are exploring this further in separate work).
We generated data under our ECHS simulation scenario with moderate noise for our propensity variable $X_1$.
Figure~\ref{fig:num-slice-impact} shows the results.
We find that as the number of slices increases, there is a slight bias in the estimated intervals, giving intervals that are on average up to 8\% smaller than the true intervals for $K = 30$ (see the mean width being lower than oracle width for larger $K$).

That being said, the bootstrap adjusted bounds contained the true parameters 97\% of the time; they were highly conservative.
Without bootstrap adjustment, however, the LQC bounds had only 50\% coverage (due to the parameters being near the extreme end of the interval).

The uncertainty in bound estimation did not increase as $K$ increased. 
The true standard errors and average estimated standard errors had no real relationship with $K$ for $K > 1$.
This shows that averaging many imperfectly estimated slices gives generally good stability in estimation.
This can be seen by how the average width of the bootstrap adjusted bounds tends to be a constant amount larger than the initially estimated bounds.

\end{document}